\documentclass[sigconf]{acmart}

\usepackage[labelformat=simple]{subcaption}
\usepackage{tabularx}

\usepackage{booktabs}  

\usepackage{graphicx}
\usepackage{bm}
\usepackage{url}
\usepackage{balance}
\usepackage{colortbl}
\usepackage{xcolor}

\usepackage{amsthm}
\usepackage{bbm}  
\usepackage{multirow}

\usepackage[linesnumbered,ruled,vlined]{algorithm2e}
\SetKwInput{KwInput}{Input}                
\SetKwInput{KwOutput}{Output}   
\usepackage{enumitem}
\usepackage{diagbox}
\usepackage{hyperref}
\usepackage{ulem}
\definecolor{TITLE_COLOR}{HTML}{f0f0f0}

\newcommand{\ie}[0]{\textit{i.e.}}
\newcommand{\eg}[0]{\textit{e.g.}}

\copyrightyear{2024}
\acmYear{2024}
\setcopyright{acmlicensed}\acmConference[KDD '24]{Proceedings of the 30th
ACM SIGKDD Conference on Knowledge Discovery and Data Mining}{August
25--29, 2024}{Barcelona, Spain}
\acmBooktitle{Proceedings of the 30th ACM SIGKDD Conference on Knowledge
Discovery and Data Mining (KDD '24), August 25--29, 2024, Barcelona, Spain}
\acmDOI{10.1145/3637528.3671749}
\acmISBN{979-8-4007-0490-1/24/08}

\def\model{\textsc{ProCom}}

\begin{document}

\author{Xixi Wu}
\author{Kaiyu Xiong}
  \affiliation{%
    \institution{Shanghai Key Laboratory of Data Science, School of Computer Science, Fudan University}
    \city{Shanghai}
    \country{China}
  }
   \email{xxwu@se.cuhk.edu.hk}
 \email{kyxiong22@m.fudan.edu.cn}

\author{Yun Xiong}\authornote{Corresponding author}
  \affiliation{%
    \institution{Shanghai Key Laboratory of Data Science, School of Computer Science, Fudan University}
    \city{Shanghai}
    \country{China}
  }
  \email{yunx@fudan.edu.cn}

\author{Xiaoxin He}
 \affiliation{
     \institution{National University of Singapore}
     \country{Singapore}
 }
  \email{xiaoxin@comp.nus.edu.sg}

\author{Yao Zhang}
  \affiliation{%
     \institution{Shanghai Key Laboratory of Data Science, School of Computer Science, Fudan University}
      \city{Shanghai}
    \country{China}
  }
  \email{yaozhang@fudan.edu.cn}
 
\author{Yizhu Jiao}
  \affiliation{%
    \institution{University of Illinois at Urbana-Champaign}
    \city{Urbana-Champaign}
    \state{IL}
    \country{USA}
  }
   \email{yizhuj2@illinois.edu}

   \author{Jiawei Zhang}
  \affiliation{%
     \institution{IFM Lab, Department of Computer Science, University of California, Davis}
     \city{Davis}
      \state{CA}
      \country{USA}
  }
   \email{jiawei@ifmlab.org}

\renewcommand{\shortauthors}{Xixi Wu et al.}
\title{\model: A Few-shot Targeted Community Detection Algorithm}

\begin{abstract}
Targeted community detection aims to distinguish a particular type of community in the network. This is an important task with a lot of real-world applications, \eg, identifying fraud groups in transaction networks. Traditional community detection methods fail to capture the specific features of the targeted community and detect all types of communities indiscriminately. Semi-supervised community detection algorithms, emerged as a feasible alternative, are inherently constrained by their limited adaptability and substantial reliance on a large amount of labeled data, which demands extensive domain knowledge and manual effort.

In this paper, we address the aforementioned weaknesses in targeted community detection by focusing on few-shot scenarios. We propose \textbf{\model}, a novel framework that extends the ``pre-train, prompt'' paradigm, offering a low-resource, high-efficiency, and transferable solution. Within the framework, we devise a dual-level context-aware pre-training method that fosters a deep understanding of latent communities in the network, establishing a rich knowledge foundation for downstream task. In the prompt learning stage, we reformulate the targeted community detection task into pre-training objectives, allowing the extraction of specific knowledge relevant to the targeted community to facilitate effective and efficient inference. By leveraging both the general community knowledge acquired during pre-training and the specific insights gained from the prompt communities, \model \space exhibits remarkable adaptability across different datasets. We conduct extensive experiments on five benchmarks to evaluate the \model \space framework, demonstrating its SOTA performance under few-shot scenarios, strong efficiency, and transferability across diverse datasets.

\end{abstract}



\maketitle

\section{Introduction}

Networks serve as powerful structures for modeling diverse relational information among objects across social \cite{SocialNetwork}, natural \cite{NaturalAcademic}, and academic domains \cite{AcademicNetwork}. A crucial step to understand a network is identifying and analyzing closely related subgraphs, \ie, communities \cite{communitygan}. However, sometimes there may exist various types of communities in the same network, while people may only focus on a specific type of community, \ie, the targeted community \cite{CLARE}. Formally, the research task of distinguishing such a targeted community from others is known as targeted community detection \cite{SEAL}. This is an important task with a lot of real-world applications, such as identifying fraud groups in transaction networks, and detecting social spammer circles in social networks \cite{SocialSpammer, FraudGroup}.

\begin{figure}[!t]
    \centering
    \includegraphics[width=8.5cm]{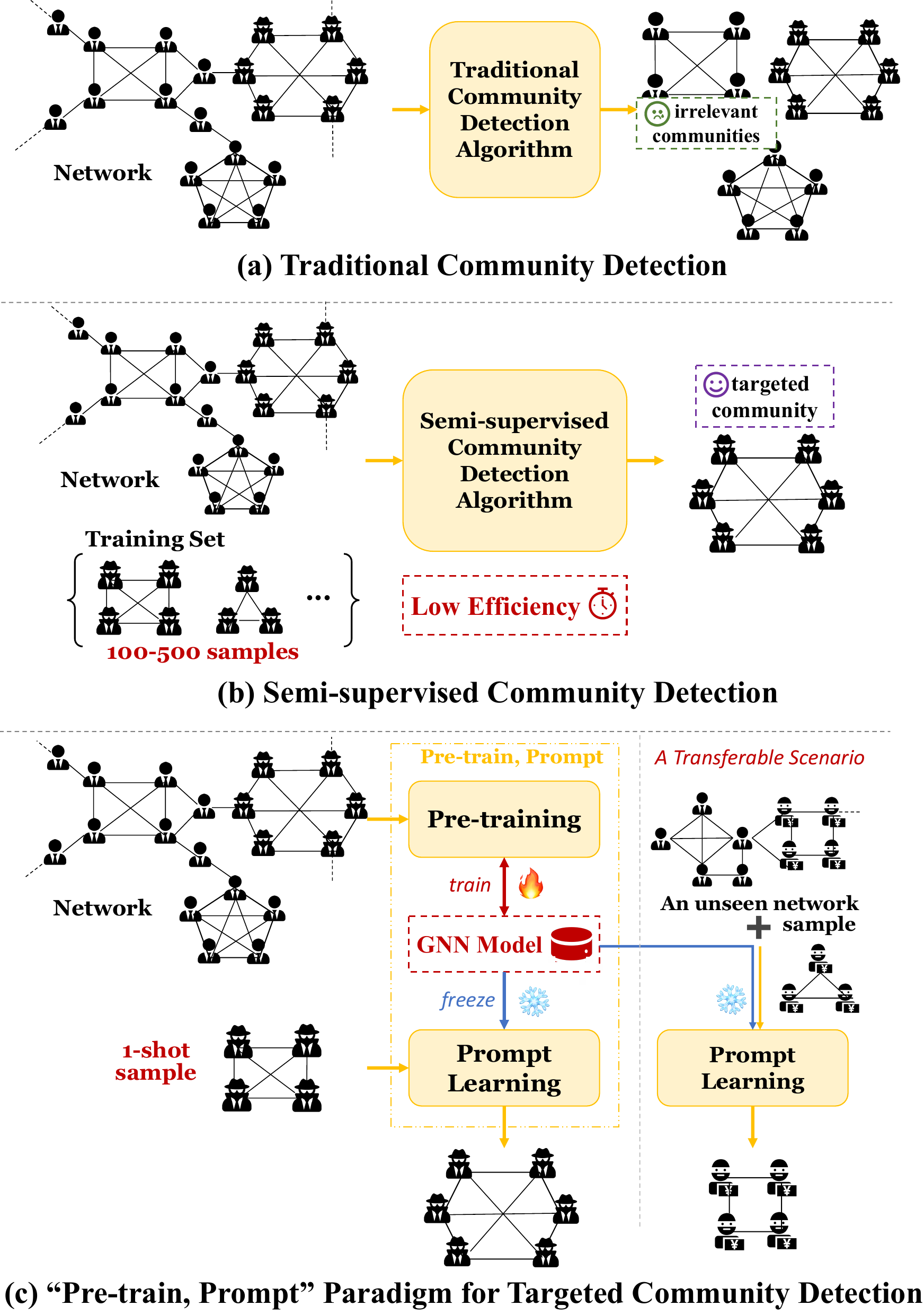}
    \vspace{-1.5em}
    \caption{A subgraph of a trading network with both normal and fraud communities. (a) Traditional community detection tends to identify both kinds of communities. (b) Semi-supervised community detection may pinpoint the remaining fraud community but requires a substantial amount of labeled data. (c) \model \space applies the ``pre-train, prompt'' paradigm to tackle the task under few-shot settings, typical in low-resource learning, efficient and transferable inference.}
    \label{fig:task_setting}
    \vspace{-1.5em}
\end{figure}

However, traditional community detection methods \cite{BigClam, CESNA, ComE, communitygan} are ill-suited for the targeted setting. This is because they exhaustively extract all types of communities from the whole network, regardless of whether they are of the same type as the targeted community or not. Taking the trading network in Figure \ref{fig:task_setting}(a) as an example, traditional community detection algorithm tends to identify not only fraud groups but also irrelevant ones, deviating from our intended goal of targeted community detection. Meanwhile, semi-supervised community detection algorithms \cite{bespoke, SEAL, CLARE} emerge as an alternative. These methods take some labeled targeted communities as input and aim to identify potential similar communities within the network. Nevertheless, a significant drawback of these approaches is their heavy reliance on a large amount of labeled data (100-500 labeled instances \cite{CLARE, SEAL}). Labeling such a substantial number of instances is laborious and requires extensive domain knowledge, making it impractical in real-world scenarios. Moreover, semi-supervised methods have limited adaptability as they necessitate gathering new relevant unlabeled data and undergoing an exhaustive  re-training process to adapt to a new targeted community detection task.

To overcome these weaknesses, the adoption of the \textit{few-shot} setting for targeted community detection emerges as a promising panacea. This setting empowers models to learn under low-resource conditions, requiring only a limited number of labeled samples. A concrete practice that aligns with this setting is the ``pre-train, prompt'' two-phase paradigm \cite{NLPprompt}. Originated from the Natural Language Processing (NLP) domain \cite{NLPprompt2, NLPPromptSurvey}, this paradigm aims to reformulate downstream task into pretexts, facilitating efficient inference with minimal downstream supervision \cite{AllinOne}. In the context of targeted community detection, the ``pre-train, prompt'' paradigm offers several advantages: (1) \textbf{Leveraging Pre-trained Models for Generalization}. During the pre-training stage, the model can gain a comprehensive understanding of various community structures and characteristics, establishing a rich knowledge foundation for downstream tasks. (2) \textbf{Using Prompt for Targeted Community Specification}. By introducing few-shot samples from the targeted community, the prompt learning stage can extract preserved knowledge specific to the targeted community type, enabling the identification of remaining targeted communities.  (3) \textbf{Adaptability}. Since the pre-trained model already incorporates general community knowledge, adapting to a new targeted community requires only a few relevant samples, resulting in minimal tuning burden and downstream supervision for seamless adaptation.

Inspired by the above insights, we propose a novel framework, \textbf{\model}, that extends the ``pre-train, \textbf{\underline{pro}}mpt'' paradigm to the targeted \textbf{\underline{com}}munity detection task, providing a low-resource, high-efficiency, and transferable solution. It incorporates a Dual-level Context-aware Pre-training method and a Targeted Community-guided Prompting Mechanism. Specifically, \model's pre-training strategy is uniquely designed to understand the latent communities in the network, equipping the model with a rich knowledge foundation for downstream tasks. The dual-level pre-training objectives involve node-to-context proximity and context distinction, capturing the underlying structures of latent communities and discriminate their distinctive features. During the prompt learning stage, we reformulate the downstream targeted community detection task into pretexts, enabling the extraction of specific knowledge to facilitate efficient inference. \model \space generates candidate communities through proximity analysis and conducts similarity matching between candidates and prompt communities. This tailored detection process aligns with pre-training objectives, ensuring both effective and efficient targeted community identification. Beyond low-resource learning and efficient inference, \model \space also exhibits transferability. Instead of relying on end-to-end frameworks, the prompting stage of \model \space can work as a plug-and-play component, swiftly adapting to any new types of targeted communities across different datasets. 

In summary, our contributions are as follows:

\begin{itemize}[leftmargin=*, topsep=2pt]
    \item We extend the ``pre-train, prompt'' paradigm for few-shot targeted community detection, addressing the heavy reliance on labeled communities. To the best of our knowledge, this is the first work that explores prompt learning specifically for community tasks.
    \item In the framework of \model, we propose a dual-level context-aware pre-training method that enables the model to acquire a rich understanding of latent communities within the network. 
    \item We further devise a targeted-community guided prompting mechanism. By aligning the downstream task with pretexts, this mechanism can extract specific knowledge relevant to the targeted community, facilitating both effective and efficient inference.
    \item Extensive experiments are conducted on diverse real-world datasets to show the SOTA performance of \model \space under few-shot settings, robustness to different numbers of prompts, strong efficiency, and transferability across different datasets.
\end{itemize}

\section{Related Works}

\subsection{Community Detection}

Community detection aims to partition graph nodes into multiple groups, where internal nodes are more similar than the external \cite{communitygan}. Traditional community detection methods can be classified into three groups: (1) Optimization-based methods \cite{Blondel_2008, Clauset_2004, 868688} reveal underlying communities by optimizing some metrics such as modularity. (2) Matrix factorization methods \cite{Li2018CommunityDI, Wang2016SemanticCI} learn latent representations for communities by decomposing adjacency matrices. (3) Generative models \cite{BigClam, CESNA} infer communities by fitting the original graph. Recently, some frameworks that combine graph representation learning and community detection have been proposed \cite{ComE, communitygan, vGraph, CGC, CommunitySurvey, CommDGI, DualStructureCom}. However, these community detection works fail to pinpoint a particular kind of community, \ie, the targeted community.

Therefore, to identify the targeted community, semi-supervised community detection methods are proposed. Bespoke \cite{bespoke} and SEAL \cite{SEAL} are seed-based methods that first locate seed nodes and then generate communities around the seed. CLARE \cite{CLARE} is the state-of-the-art semi-supervised model that proposes a novel subgraph-based inference framework. However, all these semi-supervised community detection methods necessitate a substantial number of labeled communities for effective model training.

\subsection{``Pre-train, Prompt'' on the Graph Domain}

Prompt-based tuning methods, originated from the NLP domain \cite{NLPprompt, NLPPromptSurvey}, have been widely used to facilitate the adaptation of pre-trained language models to various downstream tasks in a parameter-efficient manner. Recently, prompt learning has also emerged as a promising tool to make a pre-trained graph model seamlessly adapt to specific downstream tasks \cite{UniGPPT, AllinOne, GraphPrompt, GPPT, GPromptSurvey, Zhao2024AllIO, zi2024prog}. GPPT \cite{GPPT} performs pre-training to acquire the general structural knowledge inherent in the graph via link prediction. It then transforms the downstream node classification task into the link prediction via a hand-crafted prompting mechanism. All-in-One \cite{AllinOne}, a representative work within the graph prompt learning domain, introduces a uniform prompt design, encompassing prompt tokens, prompt structures, and inserting patterns to create a \textit{prompted} graph to enable downstream inference. Follow-up works then extend this paradigm to heterogeneous graph \cite{HetGPT}, temporal interaction graphs \cite{chen2024prompt}, text-attributed graphs \cite{Li2024ZeroGIC}, and molecular graphs \cite{wang2024ddiprompt}. 

Nevertheless, these methods are not directly applicable to the targeted community detection task due to two key limitations. Firstly, their pre-training stages lack the understanding of communities. Secondly, their prompt learning methods are specifically designed for classification tasks and rely on manipulating features to fit the provided samples. As a result, there exists a \textit{gap} between these methods and the requirements of the community-level task \cite{Qin2024PretrainAR}.
\section{Methodology}

In this section, we introduce the proposed \model \space framework. We start with the preliminaries in Section \ref{sec:pre}, and then introduce the pre-training and prompt learning stages in Sections \ref{sec:pretrain} and \ref{sec:prompt}, respectively. Notations are summarized in Table \ref{tab:notation}.

\subsection{Preliminary}\label{sec:pre}

We first give a definition of targeted community detection task and then present the \model \space pipeline to facilitate comprehension. 

\subsubsection{Targeted Community Detection} Given a graph $G=(\mathcal{V}, \mathcal{E}, \mathbf{X})$ where $\mathcal{V}$ is the set of nodes, $\mathcal{E}$ is the set of edges, and $\mathbf{X}$ denotes node feature matrix. Within the graph $G$, a community, $\mathcal{C} \subset \mathcal{V}$, denotes a subset of nodes that maintain certain desired characteristics in their edge connections or features. With the set of $m$ labeled communities $\{ \dot{\mathcal{C}}^{(i)} \}_{i=1}^m$ as training data, targeted community detection task is defined as identifying the set of other potential similar communities $\{ \hat{\mathcal{C}} \}$ within the graph. Existing semi-supervised methods \cite{SEAL, CLARE} rely on a large number of training samples by setting $m$ as 500. To reduce the reliance on labeled information, we follow the few-shot setting as fixing $m$ to a much smaller number, \eg, 10.

\subsubsection{\model \space Pipeline} We implement the \model \space framework following the ``pre-train, prompt'' two-phase paradigm \cite{AllinOne}. In the pre-training stage, we aim to acquire an understanding of the latent communities in the graph by employing carefully-designed pre-training objectives. This allows us to learn a GNN-based encoder $\text{GNN}_{\Theta}(\cdot)$. Subsequently, we obtain each node's embedding $\mathbf{z}(v) \in \mathbb{R}^d,  \forall v \in \mathcal{V}$, and we compute the embedding of a community $\mathcal{C}$ via aggregating members' representations as $\mathbf{z}(\mathcal{C}) = \sum_{v \in \mathcal{C}} \mathbf{z}(v)$. During the prompt learning phase, with $m$-shot labeled communities as prompts, our objective is to predict potential similar communities within the network while keeping $\text{GNN}_{\Theta}(\cdot)$ frozen. In this way, we leverage preserved knowledge to identify the targeted community with minimal tuning burden.

\begin{table}[!t]
    \centering
    \caption{Important Notations}
    \vspace{-1.0em}
    \begin{tabular}{c|c}
       \toprule
      \rowcolor{TITLE_COLOR} \textbf{Symbol}  & \textbf{Description} \\
       
       \midrule
       $G(\mathcal{V}, \mathcal{E}, \mathbf{X})$ & Graph \\
       $m$ & The number of prompts \\ 
       $N$ & The number of predicted communities \\ 
       \midrule

      $\text{GNN}_{\Theta}(\cdot)$ & GNN encoder parameterized by $\Theta$ \\ 
      $\text{PT}_{\Phi}(\cdot)$ & Prompting function parameterized by $\Phi$ \\
       \midrule
 
        $ \dot{\mathcal{C}}^{(i)} $ & The $i$-th ground-truth community \\
       $\mathcal{N}_v^{(k)}$  / $\mathcal{N}_v$ & ($k$)-ego net of node $v$ \\
        $\widetilde{\mathcal{N}}_v$ & Corrupted context of node $v$ \\
         $\hat{\mathcal{C}}_v = \text{PT}_{\Phi}(\mathcal{N}_v)$ & Distilled community from $\mathcal{N}_v$  \\
       \midrule 

       $\mathbf{z}(v), \mathbf{z}(\mathcal{C})$ & Embedding of node $v$, community $\mathcal{C}$ \\
       \bottomrule
       
    \end{tabular}
    \label{tab:notation}
\end{table}

\subsection{Dual-level Context-aware Pre-training}\label{sec:pretrain}
We leverage the pre-training stage to make the GNN encoder $\text{GNN}_{\Theta}(\cdot)$ aware of various structural contexts present in the network, acquiring a rich understanding of communities to benefit downstream task. Specifically, we devise dual-level pre-training objectives, \ie, node-to-context proximity and context distinction. These objectives capture the underlying structures of latent communities and discriminate their distinctive features.

\subsubsection{Node-to-Context Proximity} In the first objective, we enable the model to learn the relationships between individual nodes and the broader contexts in which they exist. Contexts often exhibit dense connections and share similar features \cite{homophily, CommunitySearch1, CommunitySearch2}, making them promising candidates for communities. By guiding the model to understand these relationships, it gains valuable insights into the community structures present in the network.

\begin{figure}[!t]
    \centering
    \includegraphics[width=8.5cm]{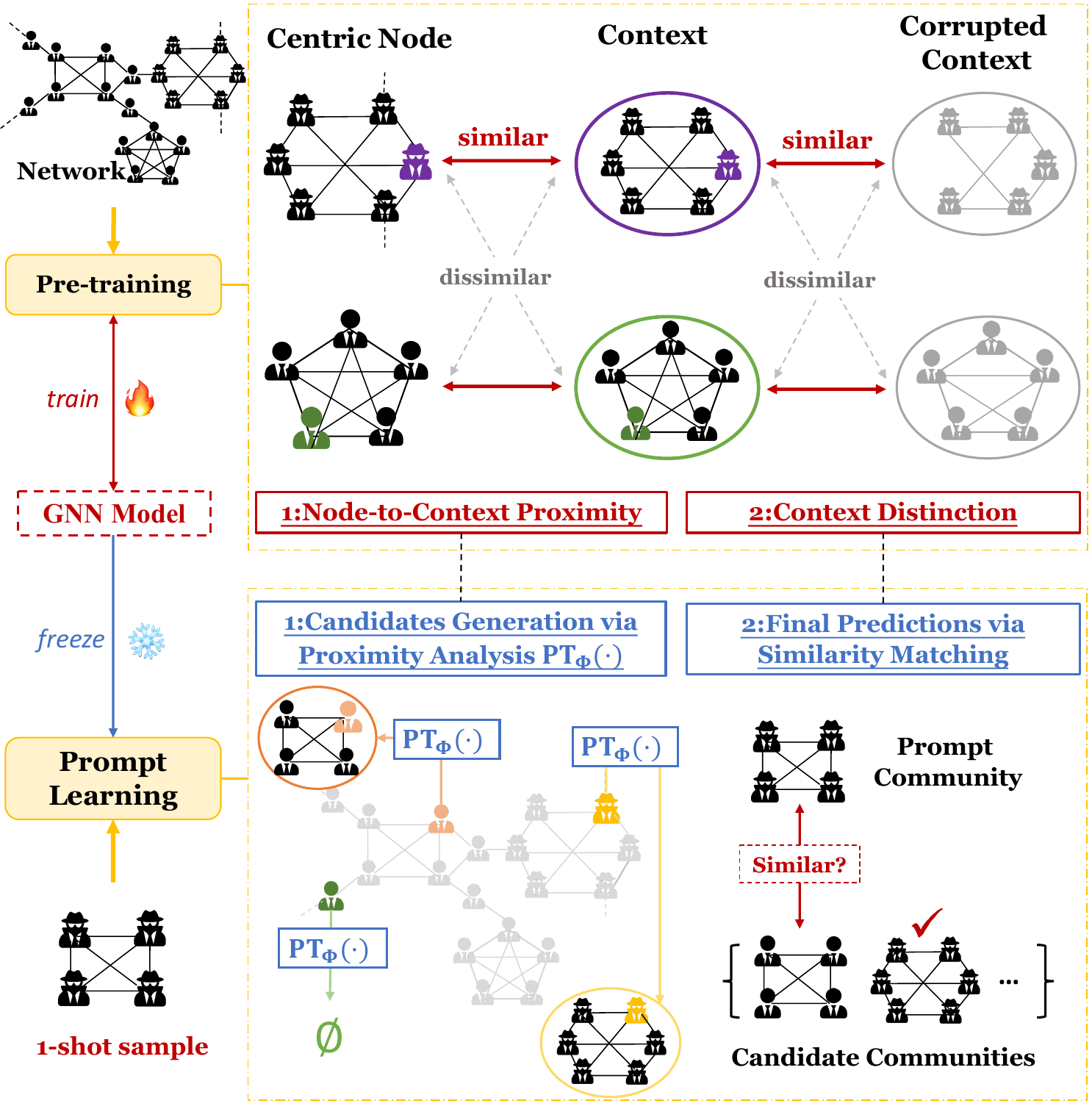}
    \vspace{-2.0em}
    \caption{Overview of \model. During the pre-training stage, we devise a dual-level pre-training method to guide the model in understanding the latent communities in the network. In the subsequent prompt learning stage, aided with few-shot samples, we reformulate the targeted community detection task into pretexts, facilitating prediction in a parameter-efficient manner.}
    \label{fig:model}
    \vspace{-1.5em}
\end{figure}

For implementation, we begin by randomly sampling a batch of nodes $\mathcal{B} \subset \mathcal{V}$. Then, for each node $v \in \mathcal{B}$, we extract its context as its $k$-ego net $\mathcal{N}_v^{(k)}$. This subgraph consists of the central node $v$ and its neighbors within $k$ hops, effectively capturing the node's surrounding environment \cite{AllinOne, ofa}. For simplicity, we omit the superscript $k$ and represent it with $\mathcal{N}_v$. After performing forward propagation on GNN, we obtain both node's and context's representations as $\mathbf{z}(v)$ and $\mathbf{z}(\mathcal{N}_v) = \sum_{r \in \mathcal{N}_v} \mathbf{z}(r)$. To encourage the alignment between each node and its surrounding context, we employ the following InfoNCE loss \cite{InfoNCE}:

\begin{equation}
 \label{eq:pretrain1}
     \mathcal{L}_{\text{n2c}}(\Theta) = \sum_{v \in \mathcal{B}} - \log \frac{ \exp \left( \text{sim}(\mathbf{z}(v), \mathbf{z}(\mathcal{N}_v))  / \tau \right)}{ \sum_{u \in \mathcal{B}} \exp \left(  \text{sim}(\mathbf{z}(v), \mathbf{z}(\mathcal{N}_u) ) / \tau \right)}  ,
\end{equation}

\noindent where ``$\text{n2c}$'' denotes ``node-to-context'', $\tau$ is a temperature hyperparameter, and the loss is parameterized by GNN's weights $\Theta$. We emphasize that for the loss, both Mean- and Sum-pooling for computing $\mathbf{z}(\mathcal{N}_v)$ result in \textbf{identical} outcomes with analysis provided in Appendix \ref{sec:proof}. By maximizing the similarity between each node and its surrounding context, the model learns to capture the intrinsic relationships between nodes and their potential affiliated communities, facilitating the understanding of community structures within the network.

\subsubsection{Context Distinction}
In the second objective, we aim to enhance the model's understanding of the characteristics of latent communities by encouraging it to differentiate between different contexts based on their unique features.

For implementation, as we have obtained node $v$'s context $\mathcal{N}_v$, we perform structural perturbation via randomly dropping nodes or edges \cite{GraphCL} to create a corrupted context $\widetilde{\mathcal{N}}_v$. Then, the optimization objective is designed to guide the alignment between representations of raw context $\mathbf{z}(\mathcal{N}_v)$ and corrupted context $\mathbf{z}(\widetilde{\mathcal{N}}_v)$:

\begin{equation}
\label{eq:pretrain2}
    \mathcal{L}_{\text{c2c}}(\Theta) = \sum_{v \in \mathcal{B}} - \log \frac{ \exp \left( \text{sim}( \mathbf{z}(\mathcal{N}_v), \mathbf{z}(\widetilde{\mathcal{N}}_v) )  / \tau \right)}{ \sum_{u \in \mathcal{B}} \exp \left(  \text{sim}( \mathbf{z}(\mathcal{N}_v), \mathbf{z}(\widetilde{\mathcal{N}}_u) ) / \tau \right)},
\end{equation}

\noindent where the subscript notation ``$\text{c2c}$'' denotes ``context-to-context''. By maximizing the similarity within a single context, we enable the model to gain insights into the distinct characteristics of latent communities present in the graph.

\subsubsection{Summary \& Discussion} The overall optimization objective combines both pre-training objectives as $ \mathcal{L}_{\text{pre-train}} =  \mathcal{L}_{\text{n2c}} + \lambda \cdot  \mathcal{L}_{\text{c2c}}$ where $\lambda \geq 0$ is a hyper-parameter that decides the importance weight of $\mathcal{L}_{\text{c2c}}$. By applying $\mathcal{L}_{\text{pre-train}}$ to learn the GNN encoder, we can obtain latent community knowledge inherent in the graph. A detailed summary of the pre-training process can be found in Algorithm \ref{algorithm:pretrain} in the Appendix. 

Furthermore, we list several widely used strategies for graph pre-training, including Node Attribute Masking \cite{GraphMAE}, Link Prediction \cite{VGAE, GraphPrompt}, Node-to-Node Consistency \cite{GRACE}, Node-to-Graph Mutual Information Maximization \cite{DGI}, and Graph-to-Graph Consistency \cite{GraphCL}. However, these approaches do not take into account the presence of communities in the network, resulting in a semantic \textit{gap} between the pre-training process and the downstream community detection task.

\subsection{Prompt Learning}\label{sec:prompt}

After the pre-training stage, the learned $\text{GNN}_{\Theta}(\cdot)$ has already acquired insights into latent communities and their distinct characteristics within the network. Then, we can obtain node-level representations by feeding the graph into this well-learned encoder, resulting in representations $\mathbf{Z}=\text{GNN}_{\Theta}(\mathbf{X}, \mathcal{E})$ for all nodes, where $\mathbf{z}(v)$ represents the embedding for node $v$.

In the prompt learning stage, our objective is to predict $N$ new communities that are similar to a given set of $m$-shot targeted communities, denoted as $\{ \dot{\mathcal{C}}^{(i)} \}_{i=1}^m$. As the pre-trained model contains rich knowledge about the targeted community, we aim to reformulate the targeted community detection task into pretexts to retrieve such knowledge. First, we generate candidate communities by performing proximity analysis between nodes and their surrounding contexts, aligning with the first pretext. This analysis helps identify potential communities based on the relationships learned by the model between nodes and their affiliated contexts. Next, we conduct similarity matching between the candidate communities and the provided prompt communities to make the final predictions, aligning with the second pretext. In this way, we can effectively leverage preserved knowledge to enhance downstream task.

\subsubsection{Candidate Community Generation} In the candidate generation phase, our objective is to generate a set of candidate communities. Communities typically exhibit stronger connections within a certain neighborhood \cite{bespoke, communitygan, CommunitySearch2}. Therefore, we aim to extract the most promising community for any given node within its context. Specifically, we introduce a prompting function $\text{PT}_{\Phi}(\cdot)$ parameterized by $\Phi$ as follows:
\begin{equation*}
    \hat{\mathcal{C}}_v = \text{PT}_{\Phi}(\mathcal{N}_v),
\end{equation*}
\noindent where $\hat{\mathcal{C}}_v$ represents the distilled community centered around node $v$, $\mathcal{N}_v$ represents the context. Note that the output of the prompting function may be an empty set, indicating that there does not exist a promising community for that particular node. In such cases, we disregard the result.

Intuitively, for each neighboring node $u$ within the context $\mathcal{N}_v$, we assign it to the distilled promising community $\hat{\mathcal{C}}_v$ only if its proximity to the central node $v$ exceeds a certain threshold value $\alpha$. As the representations of nodes, $\mathbf{z}(u)$ and $\mathbf{z}(v)$, already contain such proximity information guided by the first pretext (node-to-context proximity), we can leverage this preserved knowledge encoded in their representations. However, instead of directly utilizing pre-trained node embeddings, we introduce an additional Multi-layer Perceptron \cite{MLP} to measure proximity in a learnable manner. This allows us to extract specific knowledge relevant to the targeted community, learning the heuristics of their distinct structural patterns. The formula is given as follows:
\begin{equation*}
    \hat{ \mathcal{C}}_v = \text{PT}_{\Phi}( \mathcal{N}_v) = \{ u \in \mathcal{N}_v \; \text{and} \; \sigma(\text{MLP}_{\Phi}(\mathbf{z}(u) \| \mathbf{z}(v)) \geq \alpha \}.
\end{equation*}
\noindent Here, we implement $\text{PT}_{\Phi}(\cdot)$ using an MLP, $\mathbf{z}(u), \mathbf{z}(v)$ denote nodes $u,v$'s representations, respectively, $\sigma(\cdot)$ denotes Sigmoid function, and $\alpha \in [0, 1]$ represents a threshold parameter. By tuning the MLP with the provided prompt communities, we effectively utilize the preserved knowledge to understand the specific community structures related to the targeted community in a parameter-efficient manner.

For the optimization of prompting function $\text{PT}_{\Phi}(\cdot)$, we harness the supervision signals provided by the given prompt communities, which offer insights about structural patterns of the targeted community. Technically, we randomly select a node $v$ from a ground-truth community $\dot{\mathcal{C}}^{(i)}$ ($i=1,2,..,m$) and extract its context $\mathcal{N}_v$. Since each node $u \in \mathcal{N}_v$ can be categorized as either belonging to the ground-truth community $\dot{\mathcal{C}}^{(i)}$ or not, its status can serve as supervision for distinguishing whether a node should be included in the community. Therefore, we employ the Cross-Entropy loss to optimize $\Phi$ as follows:
\begin{equation}
\begin{split}
    \mathcal{L}_{\text{pt}}(\Phi) = \sum_{i=1}^m \sum_{v \in \dot{\mathcal{C}}^{(i)}, u \in \mathcal{N}_v}  \mathbb{I}( u, \dot{\mathcal{C}}^{(i)} ) \log \sigma \left( \text{MLP}_{\Phi}( \mathbf{z}(u) \| \mathbf{z}(v) ) \right) \\ + \left(1-  \mathbb{I}( u, \dot{\mathcal{C}}^{(i)} ) \right)  \left( 1- \log \sigma \left( \text{MLP}_{\Phi}( \mathbf{z}(u) \| \mathbf{z}(v) ) \right) \right),
\end{split}
\label{eq:prpmptune}
\end{equation}

\noindent where $\mathbb{I}(u, \mathcal{C})$ is an indicator function that returns 1 only if node $u \in \mathcal{C}$. Note that during the prompt tuning stage, we only optimize the prompting function $\text{PT}_{\Phi}(\cdot)$ while keeping $\text{GNN}_{\Theta}(\cdot)$ frozen. Additionally, due to the multiple choices for selecting node $v$ and corresponding contexts, we can generate a substantial number of supervision signals for learning $\text{PT}_{\Phi}(\cdot)$, even when given an extremely small number of $m$. The detailed prompt tuning process is summarized in Algorithm \ref{algorithm:pt} in the Appendix.

\subsubsection{Final Predictions}

In the second pretext, which focuses on context distinction, we guide the model in learning the distinct properties of different contexts. To align with this pretext and retrieve preserved knowledge, we reformulate the final predictions of the targeted community as a similarity measure between the candidate communities and prompt communities. 

After the prompt tuning process, we move into the inference stage as leveraging the well-learned prompting function $\text{PT}_{\Phi}(\cdot)$ to generate candidates. Specifically, we feed each node's ego-net $\mathcal{N}_v \; (v \in \mathcal{V}) $ to the prompting function $\text{PT}_{\Phi}(\cdot)$, obtaining a set of candidate communities $\{ \hat{\mathcal{C}}_v \}$. With both provided prompt communities $\{ \dot{\mathcal{C}}^{(i)} \}_{i=1}^m$ and candidate communities $\{ \hat{\mathcal{C}}_v \}_{v \in \mathcal{V}}$, we leverage the pre-trained encoder to obtain their respective representations: $\{ \mathbf{z}(\dot{\mathcal{C}}^{(i)})\}_{i=1}^m$ for the prompt communities and $\{ \mathbf{z}(\hat{\mathcal{C}}_v ) \}_{v \in \mathcal{V}}$ for the candidate communities. The $L_2$ distance in the latent space between these representations serves as a measure of similarity. To make the final predictions, we select the most similar candidate communities for each prompt community. For example, if we aim to predict $N$ new communities, we return the top $n = \frac{N}{m}$ most similar candidate communities for each prompt community $\dot{\mathcal{C}}^{(i)}$ \cite{CLARE}.

\begin{algorithm}[!t]
\caption{ \textbf{\model \space Pipeline}}
\label{algorithm:pipeline}
\KwInput{Graph $G(\mathcal{V},\mathcal{E},\mathbf{X})$, Prompt Communities $\{ \dot{\mathcal{C}}^{(i)} \}_{i=1}^m$, Number of Predicted Communities $N$}

\textcolor{brown}{\tcc{Pre-training}}
Perform pre-training on graph and obtain $\text{GNN}_{\Theta}(\cdot)$ based on Algorithm \ref{algorithm:pretrain}

\textcolor{brown}{\tcc{Prompt Tuning}}
Perform prompt tuning and obtain Prompting Function $\text{PT}_{\Phi}(\cdot)$ based on Algorithm \ref{algorithm:pt}

\textcolor{brown}{\tcc{Prompt-assisted Inference}}
Generate the set of candidate communities $\{ \hat{\mathcal{C}}_v \} = \{ \text{PT}_{\Phi}(\mathcal{N}_v) \}_{v \in \mathcal{V}} $

Encode prompt communities as $\mathbf{z}(\dot{\mathcal{C}}^{(i)}), \; i = 1,2,...,m$

Encode each candidate community as $\mathbf{z(}\hat{\mathcal{C}}_v) \;$

Based on $L_2$-distance in the embedding space between each $\mathbf{z}(\dot{\mathcal{C}}^{(i)})$ and $\mathbf{z(}\hat{\mathcal{C}}_v)$ as a measure of similarity, retrieve the $N$ most similar candidate communities as predicted communities

\KwOutput{Set of Final Predicted Communities $\{ \hat{\mathcal{C}} \}$}
\end{algorithm}

\subsection{Complexity Analysis} 

Due to space limitations, we move the complexity analysis to the Appendix \ref{sec:complexity}. In summary, the complexity of both pre-training and inference stages (\ie, candidate generation and similarity matching) scale linearly with the size of the graph while the prompt tuning stage (learning of $\text{PT}_{\Phi}(\cdot)$) scales linearly with the number of provided samples.

\section{Experiments}

\begin{table}[!t]
    \centering
    \caption{Statistics of datasets. The first 4 columns denote the numbers of nodes, edges, communities, and attributes, respectively. $|\overline{\mathcal{C}}|$ denotes the average community size.}

    \begin{tabular}{c|ccccc}
      \toprule
      \rowcolor{TITLE_COLOR}  \textbf{Dataset} & \textbf{\# N} & \textbf{\# E} & \textbf{\# C} & \textbf{\# A} & $|\overline{\mathcal{C}}|$ \\
       \midrule
      Facebook   &  3,622 & 72,964 & 130 & 317 & 15.6 \\ 
      Amazon     & 13,178 & 33,767 & 4,517 & - & 9.3 \\
      Livejournal & 69,860 & 911,179 & 1,000 &-  & 13.0 \\
      DBLP       & 114,095 & 466,761 & 4,559 & - &  8.4 \\ 
      Twitter    & 87,760 & 1,293,985 & 2,838 & 27,201  & 10.9 \\
      \bottomrule
    \end{tabular}
     \label{tab:dataset}
\end{table}

In this section, we present our experimental setup and empirical results. Our experiments are designed to answer the following research questions (RQs):

\begin{itemize}[leftmargin=*, topsep=2pt]
    \item \textbf{RQ1 (Overall Performance)} How does \model \space perform compared with both traditional community detection and semi-supervised community detection methods?
    \item \textbf{RQ2 (Transferability Study)} How about the generalization ability of \model? 
    \item \textbf{RQ3 (Prompt Sensitivity Study)} How do different numbers of prompt communities affect the performance of \model?
    \item \textbf{RQ4 (Efficiency Study)} How is the efficiency of \model \space compared to that of other methods?
    \item \textbf{RQ5 (Ablation Study)} How do the pre-training stage and prompt-learning stage affect the \model's performance?
\end{itemize}

\subsection{Experimental Setups}
\subsubsection{Datasets}\label{sec:aug_feature} Following previous works \cite{CLARE, SEAL, bespoke}, we choose five common real-world datasets containing overlapping communities from SNAP\footnote{http://snap.stanford.edu/data/}, including Facebook, Amazon, DBLP, Livejournal, and Twitter. Note that these datasets are partially labeled, \textit{i.e.}, most nodes do not belong to any community. Thus, we can view that there are other types of communities in the networks, and our targeted communities are the labeled ones \cite{CLARE}. Statistics of datasets are listed in Table \ref{tab:dataset}. Since Amazon, DBLP, and Livejournal do not provide nodes' attributes, we augment node features $\mathbf{x}(v) = [\text{deg}(v), \max(\text{DN}(v)), \min(\text{DN}(v)), \text{mean}(\text{DN}(v)), \text{std}(\text{DN}(v))]$ \\ where $\text{deg}(v)$ denotes the degree of node $v$, and $\text{DN}(v) = \{ \text{deg}(u) | u \in \mathcal{N}_v  \}$ as SEAL \cite{SEAL} and CLARE \cite{CLARE} do. The pre-processing details for the Livejournal dataset are explained in CLARE, while the pre-processing details for the other datasets are provided in SEAL.

\subsubsection{Baselines} To show the effectiveness of \model, we compare it with both representative traditional community detection and semi-supervised community detection methods as follows.

\noindent{\textbf{Traditional Community Detection Methods:}}
\begin{itemize}[leftmargin=*, topsep=2pt]
    \item \textbf{BigClam} \cite{BigClam}: This is a strong community detection baseline for overlapping community detection based on matrix factorization.
    \item \textbf{ComE} \cite{ComE}: This is a framework that jointly optimizes community embedding, community detection, and node embedding.
    \item \textbf{CommunityGAN}: This is a method that extends the generative model of BigClam from edge-level to motif-level.
\end{itemize}
\noindent{\textbf{Semi-supervised Community Detection Methods:}}
\begin{itemize}[leftmargin=*, topsep=2pt]
    \item \textbf{Bespoke} \cite{bespoke}: This is a semi-supervised community detection method based on structure and size information.
    \item \textbf{SEAL} \cite{SEAL}: This method aims to learn heuristics for the targeted community based on Generative Adversarial Networks.
    \item \textbf{CLARE} \cite{CLARE}: This is the state-of-the-art semi-supervised community detection algorithm that proposes a subgraph-based inference framework, including a locator and rewriter.
\end{itemize}

\begin{table}[!t]
    \centering
    \small
     \caption{Hyper-parameters in \model}

    \resizebox{0.46\textwidth}{!}{%
    \begin{tabular}{c|c|c}
      \toprule
     \rowcolor{TITLE_COLOR} \textbf{Stage}   & \textbf{Hyper-parameter} & \textbf{Value} \\
      \midrule
      \multirow{10}{*}{Pre-train}   &  Batch size $|\mathcal{B}|$ & 256 \\ 
         &  Number of epochs & 30 \\ 
         &  Learning rate & 1e-3 \\ 
         &  Basic unit of $\text{GNN}_{\Theta}(\cdot)$ & GCN \\ 
         & $k$ \& GNN layers & Search from $\{ 1, 2 \}$ \\
         &  Embedding Dimension  & 128 \\ 
         &  Temperature $\tau$ & 0.1 \\ 
         &  Remaining ratio $\rho$ for corruption & 0.85 \\ 
         &  \multirow{2}{*}{ Loss Weight $\lambda$ for $\mathcal{L}_{\text{c2c}}$} & Search from  \\
         & & $\{0.001 , 0.01, 0.1, 1\}$ \\ 

         \midrule

      \multirow{6}{*}{Prompt}   & Implementation of $\text{PT}_{\Phi}(\cdot)$ & 2 layers MLP \\ 
         & Number of epochs & 30 \\ 
         & Number of prompts $m$ & 10 \\ 
         & \multirow{2}{*}{Threshold value $\alpha$} & Search from \\
         &  & $\{ 0.1, 0.2, 0.3 \}$  \\ 
         & Learning rate & 1e-3 \\ 
         
      \toprule
    \end{tabular}
    }
    
    \label{tab:hyperparam}
\end{table}

\subsubsection{Evaluation Metrics}For networks with ground-truth communities, the most used evaluation metrics are bi-matching \textbf{F1} and \textbf{Jaccard scores} \cite{BigClam,SEAL,CLARE,bespoke}. Given $M$ ground-truth communities $\{ \dot{\mathcal{C}}^{(i)} \}$ and $N$ predicted communities $\{ \hat{\mathcal{C}}^{(j)} \}$, the scores are computed as:
\begin{equation*}
    \frac{1}{2} \left(  \frac{1}{N} \sum_{j} \max_{i} \delta ( \hat{\mathcal{C}}^{(j)}, \dot{\mathcal{C}}^{(i)} )  +  \frac{1}{M} \sum_{i} \max_{j} \delta ( \hat{\mathcal{C}}^{(j)}, \dot{\mathcal{C}}^{(i)}) \right),
\end{equation*}
\noindent where $\delta$ can be F1 or Jaccard function.

\begin{table*}[t]
    \centering
    \caption{Overall performance comparison under 10-shot setting (results in percent $\pm$ standard deviation). \\{\small ``N/A'' denotes algorithm failing to converge within 2 days. Results of traditional community detection methods are reported from CLARE and SEAL where the original papers do not provide deviations. The \colorbox{brown!30}{best} and \colorbox{brown!10}{second-best} results are highlighted with brown colors. }}
   \resizebox{0.85\textwidth}{!}{
    \begin{tabular}{c|c|ccc|ccc|cc}
       \toprule
    \multirow{2}{*}{\textbf{Metric}}   & \multirow{2}{*}{\textbf{Dataset}} & \multicolumn{3}{c|}{\textbf{Traditional Community Detection}} & \multicolumn{3}{c|}{\textbf{Semi-supervised}} & \multirow{2}{*}{\textbf{\model}} & \multirow{2}{*}{\textit{Improv.}} \\
   \cline{3-8}
         &  & BigClam & ComE & CommunityGAN & Bespoke & SEAL & CLARE &  \\
       \midrule
       \multirow{5}{*}{F1}  &  Facebook & \cellcolor{brown!10} 32.92 & 27.92 & 32.05 & $29.67_{\pm 0.85}$ & $31.10_{\pm 3.84}$ &  $28.53_{\pm 1.36}$ & \cellcolor{brown!30} $38.57_{\pm 2.02}$ &\cellcolor{brown!30} \textit{+17.2\%}\\
       
         &  Amazon   & 53.79 & 48.23 & 51.09 & $80.38_{\pm 0.64}$ & \cellcolor{brown!10}${82.26}_{\pm 4.04}$ & $78.89_{\pm 2.10}$ & \cellcolor{brown!30} ${84.36}_{\pm 0.23}$ & \cellcolor{brown!30} \textit{+2.6\%}\\ 
         
         &  Livejournal & 39.17 & N/A & 40.67 & $30.98_{\pm 1.55}$ & $42.85_{\pm 2.60}$ & \cellcolor{brown!10} ${45.38}_{\pm 4.07}$ & \cellcolor{brown!30}${54.35}_{\pm 3.04}$ & \cellcolor{brown!30}  \textit{+19.8\%}\\ 
         
         & DBLP      & 40.41 & 25.24 & N/A  & $41.55_{\pm 0.40}$ & $41.74_{\pm 6.35}$ & \cellcolor{brown!10} ${48.75}_{\pm 2.51}$ & \cellcolor{brown!30}${50.96}_{\pm 1.57}$ & \cellcolor{brown!30} \textit{+4.5\%}\\
         & Twitter  & 24.33 & 15.89 & N/A   & \cellcolor{brown!10} ${29.85}_{\pm 0.15}$ & $16.97_{\pm 1.32}$ & $20.05_{\pm 0.88}$ & \cellcolor{brown!30}${31.09}_{\pm 0.35}$ &\cellcolor{brown!30}  \textit{+4.2\%}\\

        \midrule

        \multirow{5}{*}{Jaccard}  &  Facebook & \cellcolor{brown!10}{23.47} & 18.47 & 21.04 & $20.52_{\pm 0.76}$ & $23.02_{\pm 2.98}$ & $19.64_{\pm 1.16}$ &\cellcolor{brown!30} ${28.05}_{\pm 1.85}$ & \cellcolor{brown!30} \textit{+19.5\%}\\
        
          &  Amazon   & 45.27 & 38.38 & 41.71 & $72.13_{\pm 0.55}$ & \cellcolor{brown!10}${75.44}_{\pm 4.69}$ & $68.50_{\pm 2.90}$ & \cellcolor{brown!30}${75.84}_{\pm 0.24}$ &\cellcolor{brown!30}  \textit{+0.5\%} \\
          &  Livejournal & 31.02 & N/A & 31.83 & $24.97_{\pm 1.40}$ & $35.03_{\pm 3.27}$ & \cellcolor{brown!10} ${36.38}_{\pm 3.69}$ & \cellcolor{brown!30}${44.93}_{\pm 2.96} $ &\cellcolor{brown!30}  \textit{+23.5\%} \\
         & DBLP      & 28.90 & 15.73 & N/A    & $35.42_{\pm 0.54}$ &$ 33.25_{\pm 7.05}$& \cellcolor{brown!10} ${38.30}_{\pm 2.21}$ & \cellcolor{brown!30}${39.47}_{\pm 1.64}$ &\cellcolor{brown!30}  \textit{+3.1\%} \\
         & Twitter  & 15.57 & 8.99 & N/A     & \cellcolor{brown!10} ${19.39}_{\pm 0.13}$ & $10.55_{\pm 0.98}$ & $12.52_{\pm 0.63}$ & \cellcolor{brown!30}${20.78}_{\pm 0.29}$ &\cellcolor{brown!30}  \textit{+7.2\%} \\

        \bottomrule
    \end{tabular}
    }
    \label{tab:rq1}
\end{table*}

\begin{table*}[!t]
    \centering
      \caption{Transferability study. The \colorbox{brown!10}{intra-mode} where pre-training and prompt learning are performed on the same dataset is marked with lightbrown. Transferred results that \colorbox{brown!30}{outperform intra-mode} are marked with brown. }
     \resizebox{\textwidth}{!}{
    \begin{tabular}{c|ccccc|ccccc}
      \toprule
\multirow{2}{*}{\diagbox{\textbf{Pre-train}}{\textbf{Prompt}}} & \multicolumn{5}{c|}{\textbf{F1}} & \multicolumn{5}{c}{\textbf{Jaccard}} \\
        & Facebook & Amazon & Livejournal & DBLP & Twitter & Facebook & Amazon & Livejournal & DBLP & Twitter \\
         \midrule

        Facebook & \cellcolor{brown!10} $38.57_{\pm 2.02}$ & \cellcolor{brown!30} $84.46_{\pm 0.23}$& $53.10_{\pm 2.74}$ &  \cellcolor{brown!30} $51.64_{\pm 1.32}$ &  \cellcolor{brown!30} $31.20_{\pm 0.42}$  & \cellcolor{brown!10}$28.05_{\pm 1.85}$ &  \cellcolor{brown!30} $75.95_{\pm 0.24}$ & $43.54_{\pm 2.67}$ &  \cellcolor{brown!30} $40.49_{\pm 1.52}$ &  \cellcolor{brown!30} $20.85_{\pm 0.36}$ \\
 
        Amazon & $38.04_{\pm 1.23}$ & \cellcolor{brown!10} $84.36_{\pm 0.23}$ & $53.66_{\pm 2.88}$ & $50.59_{\pm 1.30}$ & $30.15_{\pm 0.71}$ & $27.59_{\pm 1.11}$ &\cellcolor{brown!10} $75.84_{\pm 0.24}$ & $44.06_{\pm 2.86}$ & $39.27_{\pm 1.43}$ & $19.91_{\pm 0.55}$ \\ 

        Livejournal & $36.50_{\pm 1.87}$ & $83.97_{\pm 0.80}$ &\cellcolor{brown!10} $54.35_{\pm 3.04}$ &  \cellcolor{brown!30} $51.87_{\pm 1.67}$ & $28.62_{\pm 1.24}$  & $26.34_{\pm 1.66}$ & $75.46_{\pm 0.79}$ &\cellcolor{brown!10} $44.93_{\pm 2.96}$ &  \cellcolor{brown!30} $40.42_{\pm 1.78}$ & $18.80_{\pm 0.89}$ \\

        DBLP & $37.73_{\pm 1.30}$ & $84.16_{\pm 0.32}$ & $54.12_{\pm 2.60}$ & \cellcolor{brown!10} $50.96_{\pm 1.57}$ & $29.11_{\pm 1.19}$ & $27.28_{\pm 1.25}$ & $75.62_{\pm 0.31}$ & $44.56_{\pm 2.58}$ & \cellcolor{brown!10} $39.47_{\pm 1.64}$ & $19.17_{\pm 0.86}$ \\ 

        Twitter & $37.85_{\pm 2.66}$ &  \cellcolor{brown!30} $84.49_{\pm 0.26}$ & $54.03_{\pm 2.50}$ &  \cellcolor{brown!30} $51.23_{\pm 1.37}$ &\cellcolor{brown!10} $31.09_{\pm 0.35}$  & $27.59_{\pm 1.99}$ &  \cellcolor{brown!30} $75.97_{\pm 0.28}$ & $44.48_{\pm 2.55}$ &  \cellcolor{brown!30} $39.92_{\pm 1.60}$ & \cellcolor{brown!10} $20.78_{\pm 0.29}$ \\ 

         \bottomrule
    \end{tabular}
    }
    \label{tab:transfer_full}
\end{table*}

\subsubsection{Implementation Details} We implement \model \space with PyTorch and Pytorch-Geometric. We conduct all experiments on GPU machines of NVIDIA V100-32GB. Following previous works \cite{SEAL, CLARE}, for Facebook, the number of predicted communities $N$ is set as 200 while 1000 for Livejournal and 5000 for Amazon, DBLP, and Twitter, respectively. We set $m=10$ for all datasets if there is no special explanation. For all experiments, we report the averaged score and standard deviation over 10 trials. Details of \model \space hyper-parameter setting are shown in Table \ref{tab:hyperparam}. The source codes are released at \textcolor{brown}{\href{https://github.com/WxxShirley/KDD2024ProCom}{https://github.com/WxxShirley/KDD2024ProCom}}.

\subsection{Overall Performance (RQ1)}
We provide the overall performance comparison in Table \ref{tab:rq1}. For traditional community detection algorithms, we report their performance from SEAL \cite{SEAL} and CLARE \cite{CLARE} where the original papers do not provide variations. For semi-supervised algorithms and \model, we randomly select 10 communities as training data (prompts for \model) during each experiment, while the remaining communities are utilized for testing. Based on the results, we note the following key observations:
\begin{itemize}[leftmargin=*, topsep=2pt]
    \item \dotuline{Traditional community detection algorithms demonstrate inferior performance under the targeted setting}, as evidenced by their consistently suboptimal performance across all datasets.
    \item \dotuline{Semi-supervised algorithms show vulnerability when confronted with limited training data.} In the original papers of SEAL and CLARE, the number of training communities is set as 100 or 500. However, when the training data is limited to 10 communities, the performance noticeably deteriorates, falling behind even traditional community detection methods.
    \item \dotuline{\model \space outperforms all baselines in both evaluation metrics across all datasets, highlighting its superiority.} This improvement can be attributed to both the pre-training stage, which enables the acquisition of a rich understanding of latent communities in the graph, and the prompting stage, which distills specific knowledge about targeted community.
\end{itemize}

\subsection{Transferability Study (RQ2)}

In this subsection, we investigate the generalization ability of \model \space by evaluating its performance when pre-training and prompt learning are conducted on different datasets. To ensure consistent input feature dimensions across datasets and enable the transfer of pre-trained graph models, we \textbf{uniformly set node features based on their structural patterns}, as described in Section \ref{sec:aug_feature}.  The experimental results are summarized in Table \ref{tab:transfer_full}. Based on the results, we have the following key observations:
\begin{itemize}[leftmargin=*, topsep=2pt]
    \item  \dotuline{\model \space exhibits strong generalization ability.} Applying the pre-trained model to arbitrary datasets yields satisfactory performance, and, in some cases, even outperforms the intra-mode where the pre-trained model is applied on the same dataset. This can be attributed to the rich understanding of latent communities acquired during the pre-training stage, enabling the seamless adaptation of the pre-trained model to retrieve specific targeted communities from any dataset.
    
    \item Denser graphs such as Facebook and Twitter, which exhibit a greater abundance of structural patterns and latent community characteristics compared to sparser graphs like Amazon and DBLP, provide a \textbf{richer} foundation for pre-training. As a result, graph models pre-trained on these denser networks tend to achieve better performance when applied to other datasets.
\end{itemize}

\subsection{Prompt Sensitivity Study (RQ3)}

\begin{figure}[!t]
    \centering
    \vspace{-1.2em}
    \includegraphics[width=8cm]{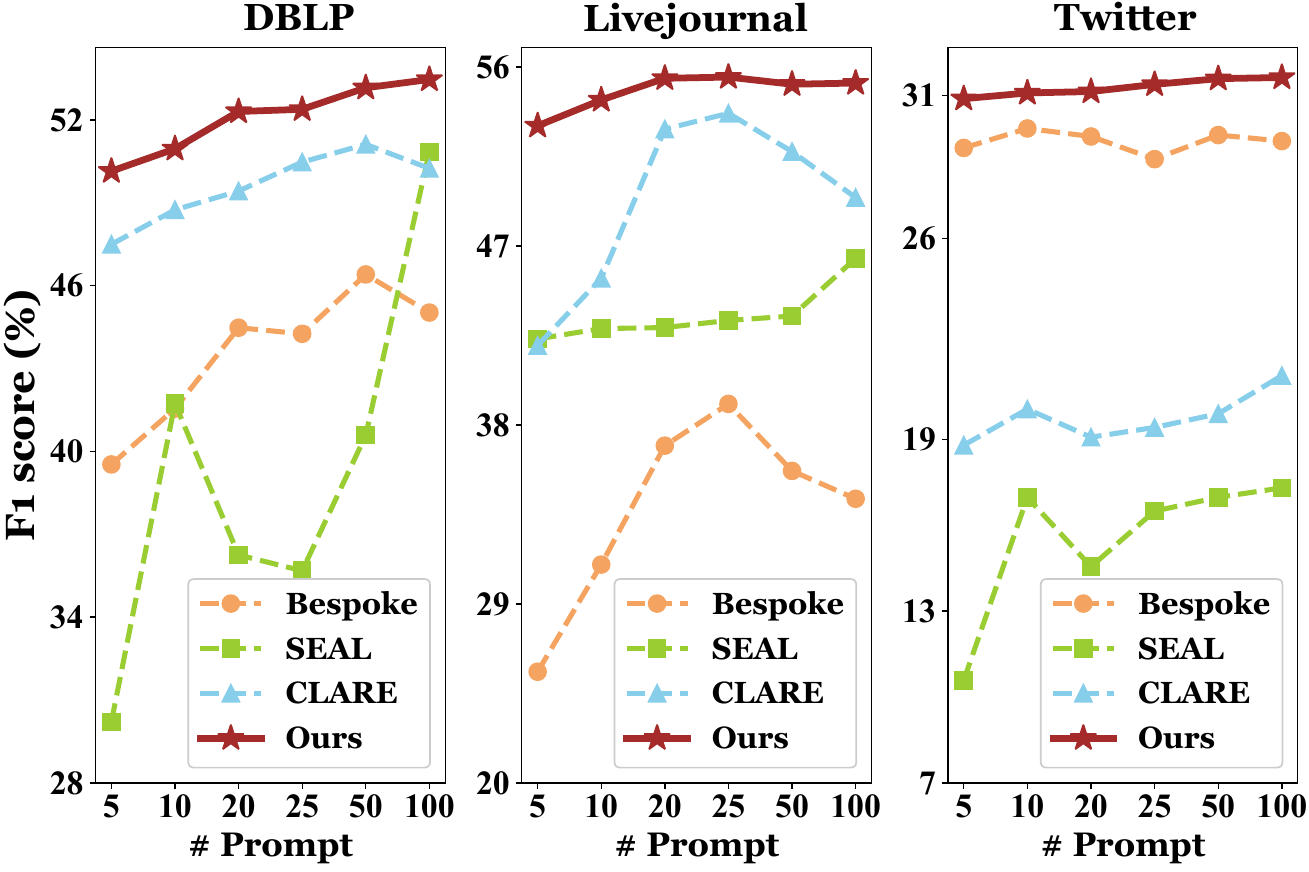}
    \caption{Prompt sensitivity study on varying numbers of prompts (training samples for semi-supervised methods). }
    \label{fig:prompt_number}
    \vspace{-1.5em}
\end{figure}

In this subsection, we delve into the sensitivity of various methods to the number of prompts. Specifically, we compare the performance of various semi-supervised community detection methods, including Bespoke \cite{bespoke}, SEAL \cite{SEAL}, and CLARE \cite{CLARE}, with that of \model \space under varying numbers of prompts. Note that for semi-supervised algorithms, these prompt communities are regarded as training communities, while for \model, they serve as prompt. We change the number of prompts from 5 to 100, and the results are presented in Figure \ref{fig:prompt_number}, where we have the following observations:
\begin{itemize}[leftmargin=*, topsep=2pt]
    \item \dotuline{Sensitivity of Semi-supervised Methods:} Semi-supervised community detection algorithms exhibit significant sensitivity to the number of training communities. When provided with a limited number of communities, their performance substantially deteriorates. 
    \item \dotuline{Robustness of \model:} (1) Even when provided with only 5 prompt communities, it consistently achieves superior performance across datasets; (2) As the number of prompt increases, the performance of \model \space tends to improve further; (3) The overall performance change within \model \space with the number of prompt ranging from 5 to 100, is relatively minor.
\end{itemize}

\subsection{Efficiency Study (RQ4)}

In this subsection, we aim to demonstrate the superior efficiency of \model \space framework. As traditional community detection methods are found to be unfeasible for the targeted task, we investigate the efficiency of semi-supervised community detection methods and compare them with \model. Specifically, we provide the total running times required for each method in Table \ref{tab:time}. For \model, we also present the required time for each stage, including pre-training, prompt tuning, and inference. It is important to note that the total running time includes not only the time required for the three stages but also the data loading and evaluation processes.

By examining the numerical values presented in the table, we can observe the remarkable efficiency of \model. Even on the largest dataset, Twitter, the total running time is a mere 446 seconds. This showcases its strong efficiency, highlighting its ability to handle the targeted setting effectively within a reasonable time frame.

\begin{table}[!t]
    \centering
    \small
     \caption{Efficiency study with numerical values of total running times. For \model, values in parentheses indicate the pre-training, prompt tuning, and inference times, respectively. ``s'', ``m'', and ``h'' denote second, minute, and hour. }
    \begin{tabular}{c|ccc|c}
     \toprule
     \rowcolor{TITLE_COLOR}  \textbf{Dataset}  & \textbf{Bespoke} & \textbf{SEAL} & \textbf{CLARE} & \textbf{\model} \\
     \midrule
       Facebook  &  4s & 2h27m & 275s & 30s (22s / 3s / 4s) \\ 
       Amazon & 110s & 1h3m & 529s & 144s (15s / 1s/ 14s) \\ 
       Livejournal & 41s & 3h35m & 832s & 260s (97s / 10s / 136s) \\ 
       DBLP & 139s & 50m & 22m & 367s (54s / 4s / 175s) \\ 
       Twitter & 126s & 2h28m & 36m & 446s (132s / 75s / 137s) \\
       \bottomrule
    \end{tabular}
    \label{tab:time}
    \vspace{-1.2em}
\end{table}

\begin{table*}[!t]
    \centering
     \caption{Ablation study on the effectiveness of our Dual-level Context-aware Pre-training. The experiments are conducted by replacing it with other graph pre-training methods within the \model \space framework.}
     \vspace{-1.0em}
    \resizebox{\textwidth}{!}{
    \begin{tabular}{c|ccccc|ccccc}
      \toprule
    \rowcolor{TITLE_COLOR}  & \multicolumn{5}{c|}{\textbf{F1}} & \multicolumn{5}{c}{\textbf{Jaccard}}  \\ 
     \rowcolor{TITLE_COLOR}   \multirow{-2}{*}{\textbf{Dataset}} & GAE  & DGI  & GraphCL & SimGRACE   & \textbf{Ours} & GAE & DGI & GraphCL & SimGRACE  & \textbf{Ours} \\
      \midrule
      Facebook & $35.64_{\pm 2.19}$ & $33.11_{\pm 3.43}$ & $29.88_{\pm 4.59}$ & $31.42_{\pm 6.22}$ & $\textbf{38.57}_{\pm 2.02}$ & $25.40_{\pm 1.75}$ & $23.47_{\pm 3.02}$ & $20.52_{\pm 3.81}$ & $21.99_{\pm 5.24}$ & $\textbf{28.05}_{\pm 1.85}$ \\ 

      Amazon & $83.70_{\pm 0.68}$ & $83.85_{\pm 0.64}$ & $84.25_{\pm 0.28}$ & $84.25_{\pm 0.24}$ & $\textbf{84.36}_{\pm 0.23}$ & $75.21_{\pm 0.59}$ & $75.30_{\pm 0.62}$ & $75.68_{\pm 0.31}$ & $75.69_{\pm 0.27}$ & $\textbf{75.84}_{\pm 0.24}$ \\ 

      Livejournal & $52.23_{\pm 4.39}$ & $52.25_{\pm 2.75}$ & $52.98_{\pm 3.09}$ & $53.44_{\pm 2.90}$ & $\textbf{54.35}_{\pm 3.04}$ & $42.88_{\pm 4.12}$ & $42.67_{\pm 2.58}$ & $43.38_{\pm 3.00}$ & $43.79_{\pm 2.86}$ & $\textbf{44.93}_{\pm 2.96}$ \\ 

      DBLP & $46.42_{\pm 2.88}$ & $47.80_{\pm 3.29}$ & $49.79_{\pm 1.89}$ &$50.68_{\pm 1.76}$ & $\textbf{50.96}_{\pm 1.57}$ & $35.33_{\pm 2.78}$ & $36.79_{\pm 2.97}$ & $38.30_{\pm 1.95}$ & $39.17_{\pm 1.82}$ & $\textbf{39.47}_{\pm 1.64}$ \\

      Twitter & $23.28_{\pm 4.29}$ & $24.62_{\pm 1.65}$ & $24.65_{\pm 3.10}$ & $28.22_{\pm 2.46}$ & $\textbf{31.09}_{\pm 0.35}$ & $15.09_{\pm 2.93}$ & $15.60_{\pm 1.22}$ & $15.68_{\pm 2.30}$ & $18.38_{\pm 1.89}$ & $\textbf{20.78}_{\pm 0.29}$ \\ 
    \bottomrule
    \end{tabular}
    }
    \label{tab:pretrain}
\end{table*}

\subsection{Ablation Study (RQ5)}

\subsubsection{Comparison with Graph Pre-training Methods}\label{sec:pretrain_experiment}
To demonstrate the effectiveness of \model's context-aware pre-training method, we compare it with several widely used strategies for pre-training GNNs. These strategies include:
\begin{itemize}[leftmargin=*, topsep=2pt]
    \item \dotuline{Node Attribute Masking} These methods \cite{VTN, GraphMAE} typically first mask node attributes and then let GNNs predict those attributes based on neighboring structures. We omit these methods from our comparison as most experimental datasets lack meaningful node attributes.
    \item \dotuline{Link Prediction} This strategy, employed by models such as \textbf{GAE} \cite{VGAE}, focuses on the reconstruction task of predicting the existence of edges between pairs of nodes.
    \item \dotuline{Node-to-Node Consistency} These methods \cite{GRACE, GCC} aim to maximize the mutual information between identical nodes under different augmented views to obtain robust node-level representations. For comparison, we use the augmentation strategies proposed in \textbf{GraphCL} \cite{GraphCL} and \textbf{SimGRACE} \cite{SimGRACE} to generate different augmented views.
    \item \dotuline{Node-to-Graph Mutual Information Maximization} These methods such as \textbf{DGI} \cite{DGI} learn the GNNs by maximizing the mutual information between a single node and the entire graph.
    \item \dotuline{Graph-to-Graph Consistency} These methods \cite{PretrainCombine2} maximize the mutual information between identical graphs under different augmented views. We omit these methods from comparison as they are applied to multi-graph datasets, whereas each of our experimental datasets consists of a single graph.
\end{itemize}

Specifically, we replace our context-aware graph pre-training method with each of the above methods while keeping all other stages consistent within the \model \space framework. The results are summarized in Table \ref{tab:pretrain}. From the results, it is evident that utilizing other graph pre-training methods consistently leads to performance degradation. This can be attributed to the fact that existing graph pre-training methods do not explicitly consider the latent communities in the network. Instead, they focus on preserving individual node features or simple contextual information. Consequently, even when provided with prompts, the retained knowledge may not be valuable in enhancing downstream predictions. \dotuline{On the contrary, our context-aware pre-training approach is specifically designed to capture the community knowledge inherent in the graph, benefiting subsequent downstream task.}

\begin{figure}[!t]
    \centering
    \includegraphics[width=7.5cm]{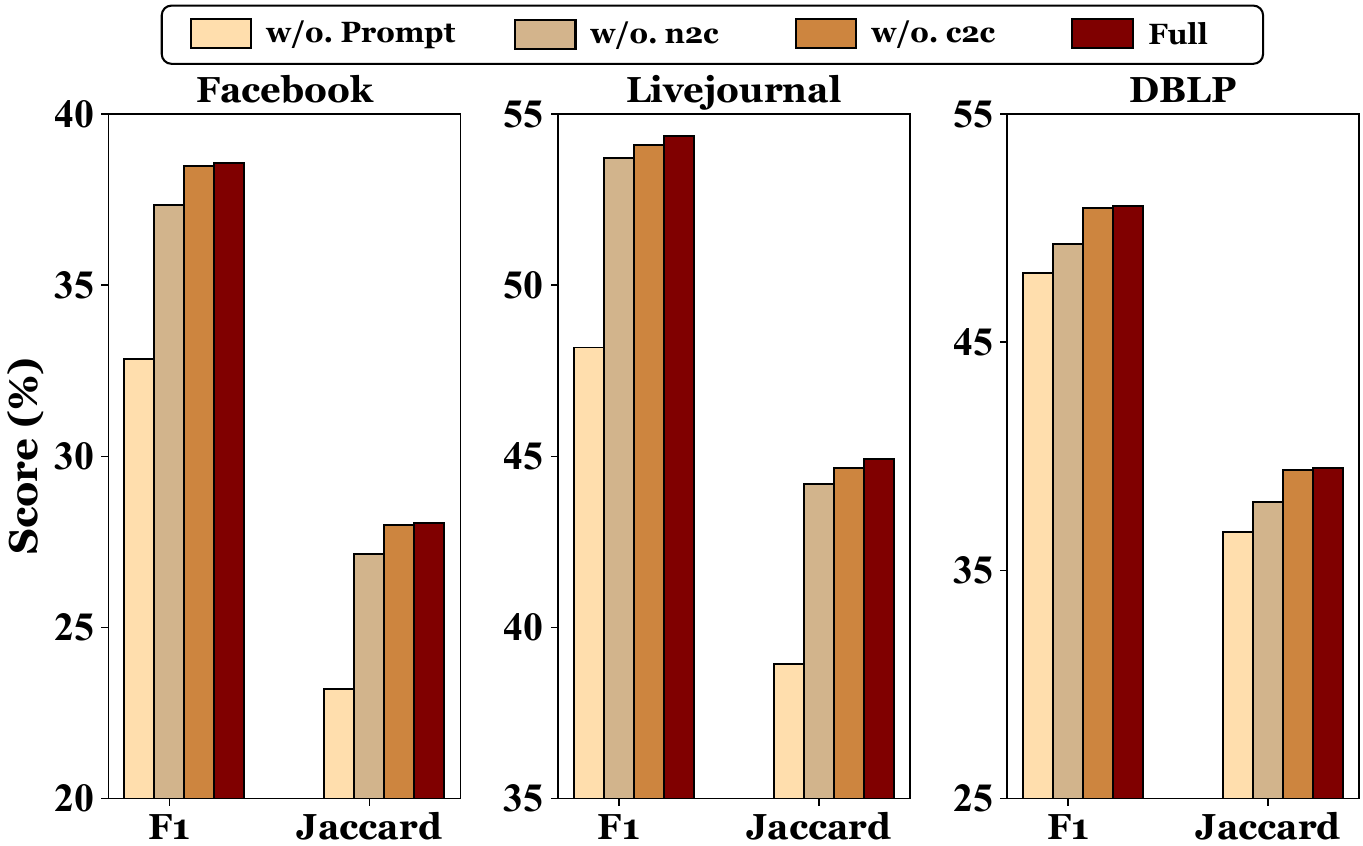}
    \caption{Ablation study on the effectiveness of both pretexts and prompt learning within the \model \space framework.}
    \label{fig:ablation}
    \vspace{-1.5em}
\end{figure}

\subsubsection{Effectiveness of Both Pretexts and Prompt Learning}\label{sec:ablation_study} 
To evaluate the effectiveness of both pre-training objectives and prompt learning, we conduct the ablation study using the following variants of the \model \space framework:
\begin{itemize}[leftmargin=*, topsep=2pt]
    \item \textbf{w/o. n2c} This variant removes the first pre-training objective, node-to-context proximity $\mathcal{L}_{\text{n2c}}$, and only utilizes the second pretext during the pre-training stage. All other stages remain consistent within the \model.
    \item \textbf{w/o. c2c} Similarly, this variant is designed to evaluate the effectiveness of the second pre-training objective, context distinction. We omit $\mathcal{L}_{\text{c2c}}$ during the pre-training stage while keeping all other designs consistent.
    \item \textbf{w/o. Prompt} This variant is designed to assess the effectiveness of prompt learning. After pre-training, we omit the prompt tuning stage and directly use the provided 10-shot communities as patterns. We consider each node's $k$-ego net as a candidate community, and predictions are made via similarity matching without any learnable process. 
\end{itemize}

The results of the ablation study are shown in Figure \ref{fig:ablation}, where we can observe that: (1) Both pre-training objectives significantly contribute to enhancing the performance of \model, as removing either of them consistently leads to poorer performance. (2) The prompt learning stage plays a crucial role within the \model, as the performance drops significantly when ``w/o. Prompt'' variant is used. This is because the assumption of $k$-ego net as a candidate community is too rigid and inflexible, resulting in inferior performance when directly matching ego-net candidates with prompt communities. In contrast, the prompt learning stage provides insights into the structural patterns specific to the targeted community, facilitating a more precise candidate generation process.

\begin{figure}[!t]
    \centering
    \includegraphics[width=7.5cm]{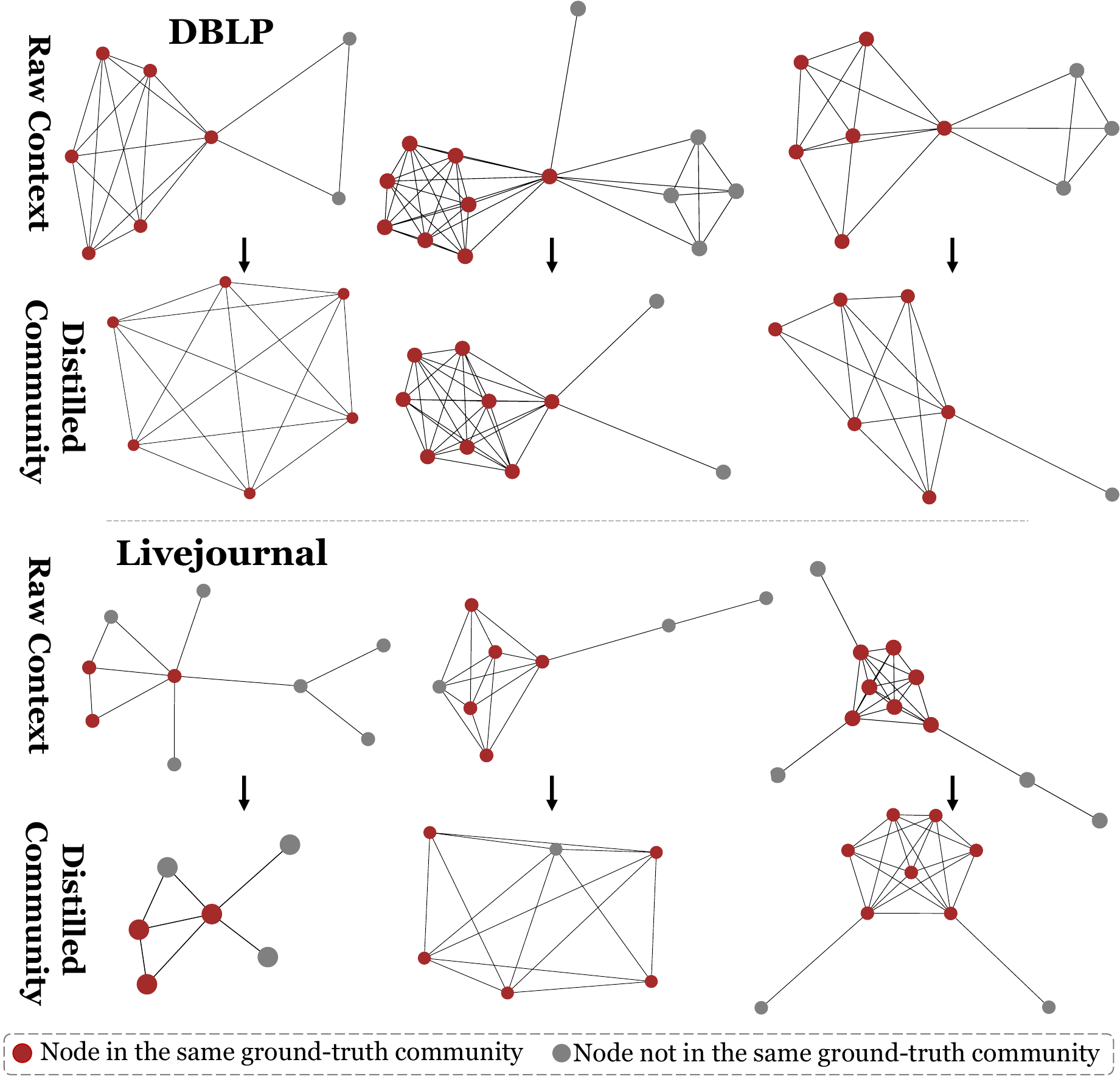}
    \caption{Case study of how prompting function works. }
    \label{fig:case}
    \vspace{-1.5em}
\end{figure}

To further illustrate the effectiveness of the prompt learning process, we conducted a case study on the DBLP and Livejournal datasets. For each predicted community, we compared its original context, \ie, ego-net, with the distilled community to assess the effectiveness of the prompting function in eliminating irrelevant nodes in Figure \ref{fig:case}. \dotuline{The results demonstrate that the prompting function effectively eliminates irrelevant nodes, resulting in distilled communities resemble the ground-truth.}

\section{Conclusion}
In this paper, we address the challenge of heavy reliance on labeled data in the targeted community detection task. To overcome this challenge, we apply the ``pre-train, prompt'' paradigm and propose \model. Our framework leverages a context-aware pre-training method to establish a rich knowledge foundation of communities present in the graph. By incorporating few-shot samples from the targeted community, the prompt learning stage extracts specific preserved knowledge, facilitating accurate inference with minimal tuning burden. \model \space showcases remarkable transferability across datasets, indicating its potential in building a graph foundational model specifically for the community detection task.

\section*{Acknowledgements}
This work is funded in part by the National Natural Science Foundation of China Projects No. U1936213. This work is also partially supported by NSF through grants IIS-1763365 and IIS-2106972.

\pagebreak
\clearpage
\newpage
\normalem
\balance{
\bibliographystyle{ACM-Reference-Format}
\bibliography{ref}


\begin{thebibliography}{52}


\ifx \showCODEN    \undefined \def \showCODEN     #1{\unskip}     \fi
\ifx \showDOI      \undefined \def \showDOI       #1{#1}\fi
\ifx \showISBNx    \undefined \def \showISBNx     #1{\unskip}     \fi
\ifx \showISBNxiii \undefined \def \showISBNxiii  #1{\unskip}     \fi
\ifx \showISSN     \undefined \def \showISSN      #1{\unskip}     \fi
\ifx \showLCCN     \undefined \def \showLCCN      #1{\unskip}     \fi
\ifx \shownote     \undefined \def \shownote      #1{#1}          \fi
\ifx \showarticletitle \undefined \def \showarticletitle #1{#1}   \fi
\ifx \showURL      \undefined \def \showURL       {\relax}        \fi
\providecommand\bibfield[2]{#2}
\providecommand\bibinfo[2]{#2}
\providecommand\natexlab[1]{#1}
\providecommand\showeprint[2][]{arXiv:#2}

\bibitem[Backstrom et~al\mbox{.}(2006)]%
        {SocialNetwork}
\bibfield{author}{\bibinfo{person}{Lars Backstrom}, \bibinfo{person}{Daniel~P. Huttenlocher}, \bibinfo{person}{Jon~M. Kleinberg}, {and} \bibinfo{person}{Xiangyang Lan}.} \bibinfo{year}{2006}\natexlab{}.
\newblock \showarticletitle{Group formation in large social networks: membership, growth, and evolution}. In \bibinfo{booktitle}{\emph{Knowledge Discovery and Data Mining}}.
\newblock


\bibitem[Bakshi et~al\mbox{.}(2018)]%
        {bespoke}
\bibfield{author}{\bibinfo{person}{Arjun Bakshi}, \bibinfo{person}{Srinivasan Parthasarathy}, {and} \bibinfo{person}{Kannan Srinivasan}.} \bibinfo{year}{2018}\natexlab{}.
\newblock \showarticletitle{Semi-Supervised Community Detection Using Structure and Size}. In \bibinfo{booktitle}{\emph{ICDM}}. \bibinfo{pages}{869--874}.
\newblock


\bibitem[Blondel et~al\mbox{.}(2008)]%
        {Blondel_2008}
\bibfield{author}{\bibinfo{person}{Vincent~D. Blondel}, \bibinfo{person}{Jean-Loup Guillaume}, \bibinfo{person}{Renaud Lambiotte}, {and} \bibinfo{person}{Etienne Lefebvre}.} \bibinfo{year}{2008}\natexlab{}.
\newblock \showarticletitle{Fast unfolding of communities in large networks}.
\newblock \bibinfo{journal}{\emph{Journal of Statistical Mechanics: Theory and Experiment}}  \bibinfo{volume}{2008} (\bibinfo{year}{2008}), \bibinfo{pages}{10008}.
\newblock


\bibitem[Brown et~al\mbox{.}(2020)]%
        {NLPprompt}
\bibfield{author}{\bibinfo{person}{Tom~B. Brown}, \bibinfo{person}{Benjamin Mann}, {and} \bibinfo{person}{Nick~Ryder et al}.} \bibinfo{year}{2020}\natexlab{}.
\newblock \showarticletitle{Language Models are Few-Shot Learners}.
\newblock \bibinfo{journal}{\emph{ArXiv}}  \bibinfo{volume}{abs/2005.14165} (\bibinfo{year}{2020}).
\newblock


\bibitem[Cavallari et~al\mbox{.}(2017)]%
        {ComE}
\bibfield{author}{\bibinfo{person}{Sandro Cavallari}, \bibinfo{person}{V. Zheng}, \bibinfo{person}{HongYun Cai}, \bibinfo{person}{K. Chang}, {and} \bibinfo{person}{E. Cambria}.} \bibinfo{year}{2017}\natexlab{}.
\newblock \showarticletitle{Learning Community Embedding with Community Detection and Node Embedding on Graphs}. In \bibinfo{booktitle}{\emph{CIKM}}.
\newblock


\bibitem[Chen et~al\mbox{.}(2024)]%
        {chen2024prompt}
\bibfield{author}{\bibinfo{person}{Xi Chen}, \bibinfo{person}{Siwei Zhang}, \bibinfo{person}{Yun Xiong}, \bibinfo{person}{Xixi Wu}, \bibinfo{person}{Jiawei Zhang}, \bibinfo{person}{Xiangguo Sun}, \bibinfo{person}{Yao Zhang}, \bibinfo{person}{Yinglong Zhao}, {and} \bibinfo{person}{Yulin Kang}.} \bibinfo{year}{2024}\natexlab{}.
\newblock \showarticletitle{Prompt Learning on Temporal Interaction Graphs}.
\newblock \bibinfo{journal}{\emph{arXiv:2402.06326}} (\bibinfo{year}{2024}).
\newblock
\showeprint[arxiv]{2402.06326}


\bibitem[Clauset et~al\mbox{.}(2004)]%
        {Clauset_2004}
\bibfield{author}{\bibinfo{person}{Aaron Clauset}, \bibinfo{person}{Mark E.~J. Newman}, {and} \bibinfo{person}{Cristopher Moore}.} \bibinfo{year}{2004}\natexlab{}.
\newblock \showarticletitle{Finding community structure in very large networks.}
\newblock \bibinfo{journal}{\emph{Physical review. E, Statistical, nonlinear, and soft matter physics}}  \bibinfo{volume}{70 6 Pt 2} (\bibinfo{year}{2004}).
\newblock


\bibitem[Fang et~al\mbox{.}(2023b)]%
        {CommunitySearch2}
\bibfield{author}{\bibinfo{person}{Shuheng Fang}, \bibinfo{person}{Kangfei Zhao}, \bibinfo{person}{Guanghua Li}, {and} \bibinfo{person}{Jeffrey~Xu Yu}.} \bibinfo{year}{2023}\natexlab{b}.
\newblock \showarticletitle{Community Search: A Meta-Learning Approach}. In \bibinfo{booktitle}{\emph{2023 IEEE 39th International Conference on Data Engineering (ICDE)}}. \bibinfo{pages}{2358--2371}.
\newblock


\bibitem[Fang et~al\mbox{.}(2023a)]%
        {UniGPPT}
\bibfield{author}{\bibinfo{person}{Taoran Fang}, \bibinfo{person}{Yunchao Zhang}, \bibinfo{person}{Yang Yang}, \bibinfo{person}{Chunping Wang}, {and} \bibinfo{person}{Lei Chen}.} \bibinfo{year}{2023}\natexlab{a}.
\newblock \showarticletitle{Universal Prompt Tuning for Graph Neural Networks}. In \bibinfo{booktitle}{\emph{Thirty-seventh Conference on Neural Information Processing Systems}}.
\newblock


\bibitem[Fang et~al\mbox{.}(2016)]%
        {CommunitySearch1}
\bibfield{author}{\bibinfo{person}{Yixiang Fang}, \bibinfo{person}{Reynold Cheng}, \bibinfo{person}{Siqiang Luo}, {and} \bibinfo{person}{Jiafeng Hu}.} \bibinfo{year}{2016}\natexlab{}.
\newblock \showarticletitle{Effective community search for large attributed graphs}.
\newblock \bibinfo{journal}{\emph{Proc. VLDB Endow.}} \bibinfo{volume}{9}, \bibinfo{number}{12} (\bibinfo{date}{aug} \bibinfo{year}{2016}), \bibinfo{pages}{1233–1244}.
\newblock
\urldef\tempurl%
\url{https://doi.org/10.14778/2994509.2994538}
\showDOI{\tempurl}


\bibitem[Hou et~al\mbox{.}(2022)]%
        {GraphMAE}
\bibfield{author}{\bibinfo{person}{Zhenyu Hou}, \bibinfo{person}{Xiao Liu}, \bibinfo{person}{Yukuo Cen}, \bibinfo{person}{Yuxiao Dong}, \bibinfo{person}{Hongxia Yang}, \bibinfo{person}{C. Wang}, {and} \bibinfo{person}{Jie Tang}.} \bibinfo{year}{2022}\natexlab{}.
\newblock \showarticletitle{GraphMAE: Self-Supervised Masked Graph Autoencoders}.
\newblock \bibinfo{journal}{\emph{Proceedings of the 28th ACM SIGKDD Conference on Knowledge Discovery and Data Mining}} (\bibinfo{year}{2022}).
\newblock


\bibitem[Hu et~al\mbox{.}(2014)]%
        {SocialSpammer}
\bibfield{author}{\bibinfo{person}{Xia Hu}, \bibinfo{person}{Jiliang Tang}, {and} \bibinfo{person}{Huan Liu}.} \bibinfo{year}{2014}\natexlab{}.
\newblock \showarticletitle{Online Social Spammer Detection}. In \bibinfo{booktitle}{\emph{AAAI Conference on Artificial Intelligence}}.
\newblock


\bibitem[Jia et~al\mbox{.}(2019)]%
        {communitygan}
\bibfield{author}{\bibinfo{person}{Yuting Jia}, \bibinfo{person}{Qinqin Zhang}, \bibinfo{person}{Weinan Zhang}, {and} \bibinfo{person}{Xinbing Wang}.} \bibinfo{year}{2019}\natexlab{}.
\newblock \showarticletitle{CommunityGAN: Community Detection with Generative Adversarial Nets}. In \bibinfo{booktitle}{\emph{WWW}}.
\newblock


\bibitem[Jin et~al\mbox{.}(2021)]%
        {CommunitySurvey}
\bibfield{author}{\bibinfo{person}{Di Jin}, \bibinfo{person}{Zhizhi Yu}, \bibinfo{person}{Pengfei Jiao}, \bibinfo{person}{Shirui Pan}, \bibinfo{person}{Philip~S. Yu}, {and} \bibinfo{person}{Weixiong Zhang}.} \bibinfo{year}{2021}\natexlab{}.
\newblock \showarticletitle{A Survey of Community Detection Approaches: From Statistical Modeling to Deep Learning}.
\newblock \bibinfo{journal}{\emph{CoRR}}  \bibinfo{volume}{abs/2101.01669} (\bibinfo{year}{2021}).
\newblock


\bibitem[Kipf and Welling(2016)]%
        {VGAE}
\bibfield{author}{\bibinfo{person}{Thomas Kipf} {and} \bibinfo{person}{Max Welling}.} \bibinfo{year}{2016}\natexlab{}.
\newblock \showarticletitle{Variational Graph Auto-Encoders}.
\newblock \bibinfo{journal}{\emph{ArXiv}}  \bibinfo{volume}{abs/1611.07308} (\bibinfo{year}{2016}).
\newblock


\bibitem[Kossinets and Watts(2009)]%
        {homophily}
\bibfield{author}{\bibinfo{person}{Gueorgi Kossinets} {and} \bibinfo{person}{Duncan~J. Watts}.} \bibinfo{year}{2009}\natexlab{}.
\newblock \showarticletitle{Origins of Homophily in an Evolving Social Network1}.
\newblock \bibinfo{journal}{\emph{Amer. J. Sociology}}  \bibinfo{volume}{115} (\bibinfo{year}{2009}), \bibinfo{pages}{405 -- 450}.
\newblock


\bibitem[Krogan et~al\mbox{.}(2006)]%
        {NaturalAcademic}
\bibfield{author}{\bibinfo{person}{Nevan~J. Krogan}, \bibinfo{person}{Gerard Cagney}, \bibinfo{person}{Haiyuan Yu}, \bibinfo{person}{Gouqing Zhong}, \bibinfo{person}{Xinghua Guo}, {and} \bibinfo{person}{et~al Alex~Ignatchenko, Joyce~Li}.} \bibinfo{year}{2006}\natexlab{}.
\newblock \showarticletitle{Global landscape of protein complexes in the yeast Saccharomyces cerevisiae}.
\newblock \bibinfo{journal}{\emph{Nature}}  \bibinfo{volume}{440} (\bibinfo{year}{2006}), \bibinfo{pages}{637--643}.
\newblock


\bibitem[Lester et~al\mbox{.}(2021)]%
        {NLPprompt2}
\bibfield{author}{\bibinfo{person}{Brian Lester}, \bibinfo{person}{Rami Al-Rfou}, {and} \bibinfo{person}{Noah Constant}.} \bibinfo{year}{2021}\natexlab{}.
\newblock \showarticletitle{The Power of Scale for Parameter-Efficient Prompt Tuning}. In \bibinfo{booktitle}{\emph{Conference on Empirical Methods in Natural Language Processing}}.
\newblock


\bibitem[Li et~al\mbox{.}(2018)]%
        {Li2018CommunityDI}
\bibfield{author}{\bibinfo{person}{Ye Li}, \bibinfo{person}{Chaofeng Sha}, \bibinfo{person}{Xin Huang}, {and} \bibinfo{person}{Yanchun Zhang}.} \bibinfo{year}{2018}\natexlab{}.
\newblock \showarticletitle{Community Detection in Attributed Graphs: An Embedding Approach}. In \bibinfo{booktitle}{\emph{AAAI}}.
\newblock


\bibitem[Li et~al\mbox{.}(2024)]%
        {Li2024ZeroGIC}
\bibfield{author}{\bibinfo{person}{Yuhan Li}, \bibinfo{person}{Peisong Wang}, \bibinfo{person}{Zhixun Li}, \bibinfo{person}{Jeffrey~Xu Yu}, {and} \bibinfo{person}{Jia Li}.} \bibinfo{year}{2024}\natexlab{}.
\newblock \showarticletitle{ZeroG: Investigating Cross-dataset Zero-shot Transferability in Graphs}.
\newblock \bibinfo{journal}{\emph{arXiv:2402.11235}} (\bibinfo{year}{2024}).
\newblock
\showeprint[arxiv]{2402.11235}


\bibitem[Liu et~al\mbox{.}(2024)]%
        {ofa}
\bibfield{author}{\bibinfo{person}{Hao Liu}, \bibinfo{person}{Jiarui Feng}, \bibinfo{person}{Lecheng Kong}, \bibinfo{person}{Ningyue Liang}, \bibinfo{person}{Dacheng Tao}, \bibinfo{person}{Yixin Chen}, {and} \bibinfo{person}{Muhan Zhang}.} \bibinfo{year}{2024}\natexlab{}.
\newblock \showarticletitle{One for All: Towards Training One Graph Model for All Classification Tasks}. In \bibinfo{booktitle}{\emph{International Conference on Learning Representations (ICLR)}}.
\newblock


\bibitem[Liu et~al\mbox{.}(2023b)]%
        {NLPPromptSurvey}
\bibfield{author}{\bibinfo{person}{Pengfei Liu}, \bibinfo{person}{Weizhe Yuan}, \bibinfo{person}{Jinlan Fu}, \bibinfo{person}{Zhengbao Jiang}, \bibinfo{person}{Hiroaki Hayashi}, {and} \bibinfo{person}{Graham Neubig}.} \bibinfo{year}{2023}\natexlab{b}.
\newblock \showarticletitle{Pre-train, Prompt, and Predict: A Systematic Survey of Prompting Methods in Natural Language Processing}.
\newblock \bibinfo{journal}{\emph{ACM Comput. Surv.}}  \bibinfo{volume}{55} (\bibinfo{year}{2023}), \bibinfo{numpages}{35}~pages.
\newblock


\bibitem[Liu et~al\mbox{.}(2023a)]%
        {GraphPrompt}
\bibfield{author}{\bibinfo{person}{Zemin Liu}, \bibinfo{person}{Xingtong Yu}, \bibinfo{person}{Yuan Fang}, {and} \bibinfo{person}{Xinming Zhang}.} \bibinfo{year}{2023}\natexlab{a}.
\newblock \showarticletitle{GraphPrompt: Unifying Pre-Training and Downstream Tasks for Graph Neural Networks}.
\newblock \bibinfo{journal}{\emph{Proceedings of the ACM Web Conference 2023}} (\bibinfo{year}{2023}).
\newblock


\bibitem[Ma et~al\mbox{.}(2024)]%
        {HetGPT}
\bibfield{author}{\bibinfo{person}{Yihong Ma}, \bibinfo{person}{Ning Yan}, \bibinfo{person}{Jiayu Li}, \bibinfo{person}{Masood Mortazavi}, {and} \bibinfo{person}{Nitesh~V. Chawla}.} \bibinfo{year}{2024}\natexlab{}.
\newblock \showarticletitle{HetGPT: Harnessing the Power of Prompt Tuning in Pre-Trained Heterogeneous Graph Neural Networks}. In \bibinfo{booktitle}{\emph{Proceedings of the ACM Web Conference 2024}}.
\newblock


\bibitem[Park et~al\mbox{.}(2022)]%
        {CGC}
\bibfield{author}{\bibinfo{person}{Namyong Park}, \bibinfo{person}{Ryan Rossi}, \bibinfo{person}{Eunyee Koh}, \bibinfo{person}{Iftikhar~Ahamath Burhanuddin}, \bibinfo{person}{Sungchul Kim}, \bibinfo{person}{Fan Du}, \bibinfo{person}{Nesreen Ahmed}, {and} \bibinfo{person}{Christos Faloutsos}.} \bibinfo{year}{2022}\natexlab{}.
\newblock \showarticletitle{CGC: Contrastive Graph Clustering ForCommunity Detection and Tracking}. In \bibinfo{booktitle}{\emph{Proceedings of the ACM Web Conference 2022}}. \bibinfo{pages}{1115–1126}.
\newblock


\bibitem[Perozzi et~al\mbox{.}(2014)]%
        {AcademicNetwork}
\bibfield{author}{\bibinfo{person}{Bryan Perozzi}, \bibinfo{person}{Leman Akoglu}, \bibinfo{person}{Patricia~Iglesias S{\'a}nchez}, {and} \bibinfo{person}{Emmanuel M{\"u}ller}.} \bibinfo{year}{2014}\natexlab{}.
\newblock \showarticletitle{Focused clustering and outlier detection in large attributed graphs}.
\newblock \bibinfo{journal}{\emph{Proceedings of the 20th ACM SIGKDD international conference on Knowledge discovery and data mining}} (\bibinfo{year}{2014}).
\newblock


\bibitem[Qin et~al\mbox{.}(2024)]%
        {Qin2024PretrainAR}
\bibfield{author}{\bibinfo{person}{Meng Qin}, \bibinfo{person}{Chaorui Zhang}, \bibinfo{person}{Yu Gao}, \bibinfo{person}{Weixi Zhang}, {and} \bibinfo{person}{Dit-Yan Yeung}.} \bibinfo{year}{2024}\natexlab{}.
\newblock \showarticletitle{Pre-train and Refine: Towards Higher Efficiency in K-Agnostic Community Detection without Quality Degradation}.
\newblock \bibinfo{journal}{\emph{ArXiv}}  \bibinfo{volume}{abs/2405.20277} (\bibinfo{year}{2024}).
\newblock


\bibitem[Qiu et~al\mbox{.}(2020)]%
        {GCC}
\bibfield{author}{\bibinfo{person}{Jiezhong Qiu}, \bibinfo{person}{Qibin Chen}, \bibinfo{person}{Yuxiao Dong}, \bibinfo{person}{Jing Zhang}, \bibinfo{person}{Hongxia Yang}, \bibinfo{person}{Ming Ding}, \bibinfo{person}{Kuansan Wang}, {and} \bibinfo{person}{Jie Tang}.} \bibinfo{year}{2020}\natexlab{}.
\newblock \showarticletitle{GCC: Graph Contrastive Coding for Graph Neural Network Pre-Training}.
\newblock \bibinfo{journal}{\emph{Proceedings of the 26th ACM SIGKDD International Conference on Knowledge Discovery \& Data Mining}} (\bibinfo{year}{2020}).
\newblock


\bibitem[Rosenblatt(1963)]%
        {MLP}
\bibfield{author}{\bibinfo{person}{Frank Rosenblatt}.} \bibinfo{year}{1963}\natexlab{}.
\newblock \showarticletitle{PRINCIPLES OF NEURODYNAMICS. PERCEPTRONS AND THE THEORY OF BRAIN MECHANISMS}.
\newblock \bibinfo{journal}{\emph{American Journal of Psychology}}  \bibinfo{volume}{76} (\bibinfo{year}{1963}), \bibinfo{pages}{705}.
\newblock


\bibitem[Shi and Malik(1997)]%
        {868688}
\bibfield{author}{\bibinfo{person}{Jianbo Shi} {and} \bibinfo{person}{Jitendra Malik}.} \bibinfo{year}{1997}\natexlab{}.
\newblock \showarticletitle{Normalized cuts and image segmentation}. In \bibinfo{booktitle}{\emph{CVPR}}. \bibinfo{pages}{731--737}.
\newblock


\bibitem[Sun et~al\mbox{.}(2019)]%
        {vGraph}
\bibfield{author}{\bibinfo{person}{Fan-Yun Sun}, \bibinfo{person}{Meng Qu}, \bibinfo{person}{Jordan Hoffmann}, \bibinfo{person}{Chin-Wei Huang}, {and} \bibinfo{person}{Jian Tang}.} \bibinfo{year}{2019}\natexlab{}.
\newblock \showarticletitle{vGraph: A Generative Model for Joint Community Detection and Node Representation Learning}. In \bibinfo{booktitle}{\emph{NIPS}}.
\newblock


\bibitem[Sun et~al\mbox{.}(2022)]%
        {GPPT}
\bibfield{author}{\bibinfo{person}{Mingchen Sun}, \bibinfo{person}{Kaixiong Zhou}, \bibinfo{person}{Xingbo He}, \bibinfo{person}{Ying Wang}, {and} \bibinfo{person}{Xin Wang}.} \bibinfo{year}{2022}\natexlab{}.
\newblock \showarticletitle{GPPT: Graph Pre-training and Prompt Tuning to Generalize Graph Neural Networks}.
\newblock \bibinfo{journal}{\emph{Proceedings of the 28th ACM SIGKDD Conference on Knowledge Discovery and Data Mining}} (\bibinfo{year}{2022}).
\newblock


\bibitem[Sun et~al\mbox{.}(2023a)]%
        {AllinOne}
\bibfield{author}{\bibinfo{person}{Xiangguo Sun}, \bibinfo{person}{Hongtao Cheng}, \bibinfo{person}{Jia Li}, \bibinfo{person}{Bo Liu}, {and} \bibinfo{person}{Jihong Guan}.} \bibinfo{year}{2023}\natexlab{a}.
\newblock \showarticletitle{All in One: Multi-Task Prompting for Graph Neural Networks}.
\newblock \bibinfo{journal}{\emph{Proceedings of the 29th ACM SIGKDD Conference on Knowledge Discovery and Data Mining}} (\bibinfo{year}{2023}).
\newblock


\bibitem[Sun et~al\mbox{.}(2023b)]%
        {GPromptSurvey}
\bibfield{author}{\bibinfo{person}{Xiangguo Sun}, \bibinfo{person}{Jiawen Zhang}, \bibinfo{person}{Xixi Wu}, \bibinfo{person}{Hong Cheng}, \bibinfo{person}{Yun Xiong}, {and} \bibinfo{person}{Jia Li}.} \bibinfo{year}{2023}\natexlab{b}.
\newblock \showarticletitle{Graph Prompt Learning: A Comprehensive Survey and Beyond}.
\newblock \bibinfo{journal}{\emph{arXiv:2311.16534}} (\bibinfo{year}{2023}).
\newblock
\showeprint[arxiv]{2311.16534}


\bibitem[Tan et~al\mbox{.}(2023)]%
        {VTN}
\bibfield{author}{\bibinfo{person}{Zhen Tan}, \bibinfo{person}{Ruocheng Guo}, \bibinfo{person}{Kaize Ding}, {and} \bibinfo{person}{Huan Liu}.} \bibinfo{year}{2023}\natexlab{}.
\newblock \showarticletitle{Virtual Node Tuning for Few-shot Node Classification}.
\newblock \bibinfo{journal}{\emph{Proceedings of the 29th ACM SIGKDD Conference on Knowledge Discovery and Data Mining}} (\bibinfo{year}{2023}).
\newblock


\bibitem[van~den Oord et~al\mbox{.}(2018)]%
        {InfoNCE}
\bibfield{author}{\bibinfo{person}{A{\"a}ron van~den Oord}, \bibinfo{person}{Yazhe Li}, {and} \bibinfo{person}{Oriol Vinyals}.} \bibinfo{year}{2018}\natexlab{}.
\newblock \showarticletitle{Representation Learning with Contrastive Predictive Coding}.
\newblock \bibinfo{journal}{\emph{ArXiv}}  \bibinfo{volume}{abs/1807.03748} (\bibinfo{year}{2018}).
\newblock


\bibitem[Velickovic et~al\mbox{.}(2018)]%
        {DGI}
\bibfield{author}{\bibinfo{person}{Petar Velickovic}, \bibinfo{person}{William Fedus}, \bibinfo{person}{William~L. Hamilton}, \bibinfo{person}{Pietro Lio’}, \bibinfo{person}{Yoshua Bengio}, {and} \bibinfo{person}{R.~Devon Hjelm}.} \bibinfo{year}{2018}\natexlab{}.
\newblock \showarticletitle{Deep Graph Infomax}.
\newblock \bibinfo{journal}{\emph{ArXiv}}  \bibinfo{volume}{abs/1809.10341} (\bibinfo{year}{2018}).
\newblock


\bibitem[Wang et~al\mbox{.}(2016)]%
        {Wang2016SemanticCI}
\bibfield{author}{\bibinfo{person}{Xiao Wang}, \bibinfo{person}{Di Jin}, \bibinfo{person}{Xiaochun Cao}, \bibinfo{person}{Liang Yang}, {and} \bibinfo{person}{Weixiong Zhang}.} \bibinfo{year}{2016}\natexlab{}.
\newblock \showarticletitle{Semantic Community Identification in Large Attribute Networks}. In \bibinfo{booktitle}{\emph{AAAI}}.
\newblock


\bibitem[Wang et~al\mbox{.}(2023)]%
        {DualStructureCom}
\bibfield{author}{\bibinfo{person}{Yuyao Wang}, \bibinfo{person}{Jie Cao}, \bibinfo{person}{Zhan Bu}, \bibinfo{person}{Jia Wu}, {and} \bibinfo{person}{Youquan Wang}.} \bibinfo{year}{2023}\natexlab{}.
\newblock \showarticletitle{Dual Structural Consistency Preserving Community Detection on Social Networks}.
\newblock \bibinfo{journal}{\emph{IEEE Transactions on Knowledge and Data Engineering}} \bibinfo{volume}{35}, \bibinfo{number}{11} (\bibinfo{year}{2023}), \bibinfo{pages}{11301--11315}.
\newblock


\bibitem[Wang et~al\mbox{.}(2024)]%
        {wang2024ddiprompt}
\bibfield{author}{\bibinfo{person}{Yingying Wang}, \bibinfo{person}{Yun Xiong}, \bibinfo{person}{Xixi Wu}, \bibinfo{person}{Xiangguo Sun}, {and} \bibinfo{person}{Jiawei Zhang}.} \bibinfo{year}{2024}\natexlab{}.
\newblock \showarticletitle{DDIPrompt: Drug-Drug Interaction Event Prediction based on Graph Prompt Learning}.
\newblock \bibinfo{journal}{\emph{arXiv:2402.11472}} (\bibinfo{year}{2024}).
\newblock
\showeprint[arxiv]{2402.11472}


\bibitem[Wu et~al\mbox{.}(2022)]%
        {CLARE}
\bibfield{author}{\bibinfo{person}{Xixi Wu}, \bibinfo{person}{Yun Xiong}, \bibinfo{person}{Yao Zhang}, \bibinfo{person}{Yizhu Jiao}, \bibinfo{person}{Caihua Shan}, \bibinfo{person}{Yiheng Sun}, \bibinfo{person}{Yangyong Zhu}, {and} \bibinfo{person}{Philip~S. Yu}.} \bibinfo{year}{2022}\natexlab{}.
\newblock \showarticletitle{CLARE: A Semi-supervised Community Detection Algorithm}. In \bibinfo{booktitle}{\emph{Proceedings of the 28th ACM SIGKDD Conference on Knowledge Discovery and Data Mining}}.
\newblock


\bibitem[Xia et~al\mbox{.}(2022)]%
        {SimGRACE}
\bibfield{author}{\bibinfo{person}{Jun Xia}, \bibinfo{person}{Lirong Wu}, \bibinfo{person}{Jintao Chen}, \bibinfo{person}{Bozhen Hu}, {and} \bibinfo{person}{Stan~Z. Li}.} \bibinfo{year}{2022}\natexlab{}.
\newblock \showarticletitle{SimGRACE: A Simple Framework for Graph Contrastive Learning without Data Augmentation}. In \bibinfo{booktitle}{\emph{Proceedings of the ACM Web Conference 2022}}.
\newblock


\bibitem[Yang and Leskovec(2013)]%
        {BigClam}
\bibfield{author}{\bibinfo{person}{Jaewon Yang} {and} \bibinfo{person}{Jure Leskovec}.} \bibinfo{year}{2013}\natexlab{}.
\newblock \showarticletitle{Overlapping community detection at scale: a nonnegative matrix factorization approach}. In \bibinfo{booktitle}{\emph{WSDM}}. \bibinfo{pages}{587--596}.
\newblock


\bibitem[Yang et~al\mbox{.}(2013)]%
        {CESNA}
\bibfield{author}{\bibinfo{person}{Jaewon Yang}, \bibinfo{person}{Julian McAuley}, {and} \bibinfo{person}{Jure Leskovec}.} \bibinfo{year}{2013}\natexlab{}.
\newblock \showarticletitle{Community Detection in Networks with Node Attributes}. In \bibinfo{booktitle}{\emph{ICDM}}.
\newblock


\bibitem[You et~al\mbox{.}(2021)]%
        {PretrainCombine2}
\bibfield{author}{\bibinfo{person}{Yuning You}, \bibinfo{person}{Tianlong Chen}, \bibinfo{person}{Yang Shen}, {and} \bibinfo{person}{Zhangyang Wang}.} \bibinfo{year}{2021}\natexlab{}.
\newblock \showarticletitle{Graph Contrastive Learning Automated}. In \bibinfo{booktitle}{\emph{International Conference on Machine Learning}}.
\newblock


\bibitem[You et~al\mbox{.}(2020)]%
        {GraphCL}
\bibfield{author}{\bibinfo{person}{Yuning You}, \bibinfo{person}{Tianlong Chen}, \bibinfo{person}{Yongduo Sui}, \bibinfo{person}{Ting Chen}, \bibinfo{person}{Zhangyang Wang}, {and} \bibinfo{person}{Yang Shen}.} \bibinfo{year}{2020}\natexlab{}.
\newblock \showarticletitle{Graph Contrastive Learning with Augmentations}. In \bibinfo{booktitle}{\emph{NIPS}}.
\newblock


\bibitem[Yu et~al\mbox{.}(2023)]%
        {FraudGroup}
\bibfield{author}{\bibinfo{person}{Jianke Yu}, \bibinfo{person}{Hanchen Wang}, \bibinfo{person}{Xiaoyang Wang}, \bibinfo{person}{Zhao Li}, \bibinfo{person}{Lu Qin}, \bibinfo{person}{Wenjie Zhang}, \bibinfo{person}{Jian Liao}, {and} \bibinfo{person}{Ying Zhang}.} \bibinfo{year}{2023}\natexlab{}.
\newblock \showarticletitle{Group-based Fraud Detection Network on e-Commerce Platforms}. In \bibinfo{booktitle}{\emph{Proceedings of the 29th ACM SIGKDD Conference on Knowledge Discovery and Data Mining}}. \bibinfo{pages}{5463–5475}.
\newblock


\bibitem[Zhang et~al\mbox{.}(2020b)]%
        {CommDGI}
\bibfield{author}{\bibinfo{person}{Tianqi Zhang}, \bibinfo{person}{Yun Xiong}, \bibinfo{person}{Jiawei Zhang}, \bibinfo{person}{Yao Zhang}, \bibinfo{person}{Yizhu Jiao}, {and} \bibinfo{person}{Yangyong Zhu}.} \bibinfo{year}{2020}\natexlab{b}.
\newblock \showarticletitle{CommDGI: Community Detection Oriented Deep Graph Infomax}. In \bibinfo{booktitle}{\emph{Proceedings of the 29th ACM International Conference on Information \& Knowledge Management}}. \bibinfo{pages}{1843–1852}.
\newblock


\bibitem[Zhang et~al\mbox{.}(2020a)]%
        {SEAL}
\bibfield{author}{\bibinfo{person}{Yao Zhang}, \bibinfo{person}{Yun Xiong}, \bibinfo{person}{Yun Ye}, \bibinfo{person}{Tengfei Liu}, \bibinfo{person}{Weiqiang Wang}, \bibinfo{person}{Yangyong Zhu}, {and} \bibinfo{person}{Philip~S. Yu}.} \bibinfo{year}{2020}\natexlab{a}.
\newblock \showarticletitle{SEAL: Learning Heuristics for Community Detection with Generative Adversarial Networks}. In \bibinfo{booktitle}{\emph{Proceedings of the 26th ACM SIGKDD International Conference on Knowledge Discovery \& Data Mining}}.
\newblock


\bibitem[Zhao et~al\mbox{.}(2024)]%
        {Zhao2024AllIO}
\bibfield{author}{\bibinfo{person}{Haihong Zhao}, \bibinfo{person}{Aochuan Chen}, \bibinfo{person}{Xiangguo Sun}, \bibinfo{person}{Hong Cheng}, {and} \bibinfo{person}{Jia Li}.} \bibinfo{year}{2024}\natexlab{}.
\newblock \showarticletitle{All in One and One for All: A Simple yet Effective Method towards Cross-domain Graph Pretraining}.
\newblock \bibinfo{journal}{\emph{ArXiv}}  \bibinfo{volume}{abs/2402.09834} (\bibinfo{year}{2024}).
\newblock


\bibitem[Zhu et~al\mbox{.}(2020)]%
        {GRACE}
\bibfield{author}{\bibinfo{person}{Yanqiao Zhu}, \bibinfo{person}{Yichen Xu}, \bibinfo{person}{Feng Yu}, \bibinfo{person}{Q. Liu}, \bibinfo{person}{Shu Wu}, {and} \bibinfo{person}{Liang Wang}.} \bibinfo{year}{2020}\natexlab{}.
\newblock \showarticletitle{Deep Graph Contrastive Representation Learning}.
\newblock \bibinfo{journal}{\emph{ArXiv}}  \bibinfo{volume}{abs/2006.04131} (\bibinfo{year}{2020}).
\newblock


\bibitem[Zi et~al\mbox{.}(2024)]%
        {zi2024prog}
\bibfield{author}{\bibinfo{person}{Chenyi Zi}, \bibinfo{person}{Haihong Zhao}, \bibinfo{person}{Xiangguo Sun}, \bibinfo{person}{Yiqing Lin}, \bibinfo{person}{Hong Cheng}, {and} \bibinfo{person}{Jia Li}.} \bibinfo{year}{2024}\natexlab{}.
\newblock \showarticletitle{ProG: A Graph Prompt Learning Benchmark}.
\newblock \bibinfo{journal}{\emph{arXiv:2406.05346}} (\bibinfo{year}{2024}).
\newblock
\showeprint[arxiv]{2406.05346}


\end{thebibliography}
}

\pagebreak
\appendix
\section{Algorithm Details}
In this section, we show the detailed algorithm processs of pre-training and prompt tuning in Algorithms \ref{algorithm:pretrain} and \ref{algorithm:pt}, respectively.

\begin{algorithm}[!h]
\caption{\textbf{\model \space Pre-training}}
\label{algorithm:pretrain}
\KwInput{Graph $G(\mathcal{V},\mathcal{E},\mathbf{X})$}

Initialize $\text{GNN}_{\Theta}(\cdot)$

\While{not converge}
{
   Randomly sample a batch of nodes $\mathcal{B} \subset \mathcal{V}$

   \For{$v \in \mathcal{B}$}
   {
      Extract node $v$'s context $\mathcal{N}_v$, represent its corresponding subgraph as $(\mathbf{X}_v, \mathbf{A}_v)$

      Encode this subgraph as $\mathbf{Z}_v$ = $\text{GNN}_{\Theta}( \mathbf{X}_v, \mathbf{A}_v)$, retrieve node representation $\mathbf{z}(v)$ and context representation $\mathbf{z}(\mathcal{N}_v) = \text{Sum-Pooling}(\mathbf{Z}_v)$

      Compute corrupted context representation $\mathbf{z}(\widetilde{\mathcal{N}}_v) = \text{Sum-Pooling}(\text{GNN}_{\Theta}(\mathbf{X}_v, \widetilde{\mathbf{A}}_v)) $
     
   }

   Update $\Theta$ by applying gradient descent to minimize $\mathcal{L}_{\text{pre-train}}$ based on Equations \ref{eq:pretrain1} and \ref{eq:pretrain2}
 
}
\KwOutput{Pre-trained Graph Model $\text{GNN}_{\Theta}(\cdot)$}
\end{algorithm}

\begin{algorithm}[!h]
\caption{\textbf{\model \space Prompt Tuning}}
\label{algorithm:pt}
\KwInput{Graph $G(\mathcal{V},\mathcal{E},\mathbf{X})$, Pre-trained Model $\text{GNN}_{\Theta}(\cdot)$, Prompt Communities $\{ \dot{\mathcal{C}}^{(i)} \}_{i=1}^m$}

Obtain all nodes' representations as $\mathbf{Z} =  \text{GNN}_{\Theta}(\mathbf{X}, \mathcal{E})$

Initialize Prompting Function $\text{PT}_{\Phi}(\cdot)$

\While{not converge}
{
   Pick a node $v$ from a randomly sampled prompt community $\dot{\mathcal{C}}$

   Extract $v$'s ego-net $\mathcal{N}_v$, retrieve node embeddings $\{ \mathbf{z}(u) \}_{u \in \mathcal{N}_v} $ from $\mathbf{Z}$

    Update $\Phi$ by applying gradient descent to minimize $\mathcal{L}_{\text{pt}}$ based on Equation \ref{eq:prpmptune}
}
\KwOutput{Tuned Prompting Function $\text{PT}_{\Phi}(\cdot)$}
\end{algorithm}

\section{Discussion of Pooling Operations}\label{sec:proof}

We provide a proof that both Mean- and Sum-pooling methods for computing $\mathbf{z}(\mathcal{N}_v)$ result in \textbf{identical} outcomes when calculating $\mathcal{L}_{\text{pre-train}}$.

Recall that $ \mathcal{L}_{\text{n2c}} = \sum_{v \in \mathcal{B}} - \log \frac{ \exp \left( \text{sim}(\mathbf{z}(v), \mathbf{z}(\mathcal{N}_v))  / \tau \right)}{ \sum_{u \in \mathcal{B}} \exp \left(  \text{sim}(\mathbf{z}(v), \mathbf{z}(\mathcal{N}_u) ) / \tau \right)}$. Given $\text{sim}( \mathbf{a}, \mathbf{b}) = \frac{ \mathbf{a} \cdot \mathbf{b} }{ \| \mathbf{a} \| \| \mathbf{b} \| }$, if we compute $\mathbf{z}(\mathcal{N}_v)$ via Mean-Pooling as $\mathbf{z}(\mathcal{N}_v) = \frac{1}{|\mathcal{N}_v|} \sum_{k \in \mathcal{N}_v} \mathbf{z}(k)$, the factor $\frac{1}{|\mathcal{N}_v|}$ can be divided by both the numerator and the denominator in the $\text{sim}(\cdot, \cdot)$. Consequently, this is equivalent to computing $\mathbf{z}(\mathcal{N}_v)$ using Sum-pooling. Therefore, we can conclude that both Mean- and Sum-pooling operations yield the same results for this loss function. Similar reasoning can be applied to $\mathcal{L}_{\text{c2c}}$, leading to the conclusion that both Mean- and Sum-pooling operations yield the same results for $\mathcal{L}_{\text{pre-train}}$.

\section{Complexity Analysis}\label{sec:complexity}
In this section, we analyze the complexity of the \model \space framework. We discuss the complexity of each stage in detail with the graph defined as $G=(\mathcal{V}, \mathcal{E}, \mathbf{X})$.

\begin{itemize}[leftmargin=*, topsep=2pt]
    \item \textbf{Pre-training (Algorithm \ref{algorithm:pretrain})} Within each epoch, we randomly sample a batch of nodes $\mathcal{B}$ and extract their contexts to compute $\mathcal{L}_{\text{n2c}}$. The complexity of corresponding graph convolution is $\mathcal{O}(|\mathcal{E}'|Ld)$ where $|\mathcal{E}'| \ll |\mathcal{E}|$ represents the sum of edges within each context, $L=k$ denotes the number of convolutional layers, and $d$ represents the embedding dimension. Additionally, the complexity of computing $\mathcal{L}_{\text{n2c}}$ is $\mathcal{O}(|\mathcal{B}|d + |\mathcal{B}|^2d)$ and we have $|\mathcal{B}| \ll |\mathcal{V}|$. Next, we analyze the computation of $\mathcal{L}_{\text{c2c}}$. The complexity of graph convolution on the corrupted context is $\mathcal{O}(\rho |\mathcal{E}'|Ld)$ where $\rho \in (0, 1]$ represents the ratio of remaining edges. Therefore, the complexity within one pre-training epoch is $\mathcal{O}\left( (1+\rho) |\mathcal{E}'|Ld + 2(|\mathcal{B}|d + |\mathcal{B}|^2 d)  \right) \textless \mathcal{O}(|\mathcal{E}| + |\mathcal{V}|) $.

    \item \textbf{Prompt Tuning (Algorithm \ref{algorithm:pt})} Before prompt tuning, we first compute the embeddings of all nodes based on the pre-trained graph model (Line 1), resulting in a complexity of $\mathcal{O}(|\mathcal{E}|Ld)$. We use $\bar{|\mathcal{N}|} \ll |\mathcal{V}|$ to denote the average size of a node's context, \ie, ego-net. Within each epoch of prompt tuning (Lines 4-6), the complexity is $\mathcal{O}(m\bar{|\mathcal{N}|}L'd) \ll \mathcal{O}(|\mathcal{V}|)$ where $L'$ denotes the number of layers within $\text{PT}_{\Phi}(\cdot)$ and $m$ refers to the number of prompt communities. 

    \item \textbf{Inference (Lines 6-9 in Algorithm \ref{algorithm:pipeline})} The inference process includes two steps, \ie, candidate generation and similarity matching. For candidate generation, we extract the context of each node and feed it to the well-tuned prompting function (Line 6), resulting in a complexity of $\mathcal{O}(|\mathcal{V}||\bar{\mathcal{N}}|L'd)$ where $|\bar{\mathcal{N}}|$ still represents the average size of a node's context. Before similarity matching, we encode both candidate and prompt communities (Lines 7-8) with a complexity that does not exceed $\mathcal{O}(|\mathcal{E}|Ld)$. And the complexity of similarity matching ($L_2$ Distance) between candidates and prompts is less than $\mathcal{O}(m |\mathcal{V}|d)$ since the number of distilled candidates is less than the number of nodes $|\mathcal{V}|$ in the graph. Therefore, the overall complexity of the inference stage is smaller than $\mathcal{O}(|\mathcal{V}||\bar{\mathcal{N}}|L'd + |\mathcal{E}|Ld + m|\mathcal{V}|d) \textless \mathcal{O}(|\mathcal{E}| + |\mathcal{V}|)$.
    
\end{itemize}

\dotuline{Based on the above analysis, the complexity of both pre-training and inference stages within \model \space scales linearly with the number of nodes and edges in the graph. The complexity of prompt tuning scales linearly with the number of provided samples, demonstrating the efficiency.}

\end{document}